\documentclass[12pt,preprint]{aastex}

\renewcommand{\deg}{\mbox{$^{\circ}$}}

\def\farcm{\hbox{$.\mkern-4mu^\prime$}}
\def\farcs{\hbox{$.\!\!^{\prime\prime}$}}
\def\fdg{\hbox{$.\!\!^\circ$}}

\newcommand{\omc}{\mbox{$\omega$ Cen~}} 
\newcommand{\omcp}{\mbox{$\omega$ Cen}}

\shorttitle{Star Counts in {\boldmath $\omega$}\,Centauri}
\shortauthors{Castellani, et al.}

\begin{document}



\title{Star Counts in the Globular Cluster {\boldmath $\omega$}\,Centauri. I. 
Bright Stellar Components\altaffilmark{1,2}}

\author{
V. Castellani\altaffilmark{3},
A. Calamida\altaffilmark{3,4},
G. Bono\altaffilmark{3,5},
P. B. Stetson\altaffilmark{6,14,15},
L. M. Freyhammer\altaffilmark{7},
S. Degl'Innocenti\altaffilmark{8,9}
P. Prada Moroni\altaffilmark{8,9}
M. Monelli\altaffilmark{10},
C. E. Corsi\altaffilmark{3},
M. Nonino\altaffilmark{11},
R. Buonanno\altaffilmark{4},
F. Caputo\altaffilmark{3},
M. Castellani\altaffilmark{3},
M. Dall'Ora \altaffilmark{12},
M. Del Principe\altaffilmark{13},
I. Ferraro\altaffilmark{3},
G. Iannicola\altaffilmark{3},
A. M. Piersimoni\altaffilmark{13},
L. Pulone\altaffilmark{3}, 
C. Vuerli\altaffilmark{11},
}

\altaffiltext{1}{During the revision of this manuscript Vittorio Castellani 
passed away on May 19, 2006. His suggestions, ideas, and personality will be
greatly missed.}

\altaffiltext{2}{Based on data obtained from the ESO Science Archive Facility 
and the Hubble Space Telescope Archive Facility.}

\altaffiltext{3}{INAF-Osservatorio Astronomico di Roma, Via Frascati 33, 00040, 
Monte Porzio Catone, Italy; bono@mporzio.astro.it, caputo@mporzio.astro.it, 
m.castellani@mporzio.astro.it, corsi@mporzio.astro.it, giacinto@mporzio.astro.it, 
ferraro@mporzio.astro.it, pulone@mporzio.astro.it}

\altaffiltext{4}{Universita' di Roma Tor Vergata, Via della Ricerca Scientifica 1,
00133 Rome, Italy, buonanno@mporzio.astro.it, calamida@mporzio.astro.it}

\altaffiltext{5}{European Southern Observatory, Karl-Schwarzschild-Str. 2, 
D-85748 Garching bei Munchen, Germany}

\altaffiltext{6}{
Dominion Astrophysical Observatory, Herzberg Institute of Astrophysics,
National Research Council, 5071 West Saanich Road, Victoria, BC V9E~2E7,
Canada; Peter.Stetson@nrc-cnrc.gc.ca}

\altaffiltext{7}{Centre for Astrophysics, University of Central Lancashire, 
Preston PR1 2HE; lmfreyhammer@uclan.ac.uk}

\altaffiltext{8}{Dipartimento di Fisica "E. Fermi", Univ. Pisa, 
Largo B. Pontecorvo 3, 56127 Pisa, Italy; prada@df.unipi.it, scilla@df.unipi.it}

\altaffiltext{9}{INFN, Sez. Pisa, via Largo B. Pontecorvo 2, 56127 Pisa, Italy}

\altaffiltext{10}{IAC - Instituto de Astrofisica de Canarias, Calle Via Lactea,
E38200 La Laguna, Tenerife, Spain; monelli@iac.es}

\altaffiltext{11}{INAF-Osservatorio Astronomico di Trieste, via G.B. Tiepolo 11,
40131 Trieste, Italy; nonino@ts.astro.it, vuerli@ts.astro.it} 

\altaffiltext{12}{INAF - Osservatorio Astronomico di Capodimonte,
via Moiariello 16, 80131 Napoli; dallora@na.astro.it}

\altaffiltext{13}{INAF-Osservatorio Astronomico di Collurania, via M. Maggini, 
64100 Teramo, Italy; milena@te.astro.it, piersimoni@te.astro.it}

\altaffiltext{14}{Guest User, Canadian Astronomy Data Centre, which is operated
by the Herzberg Institute of Astrophysics, National Research Council of Canada.}

\altaffiltext{15}{Guest Investigator of the UK Astronomy Data Centre.}

\begin{abstract}
We present an extensive photometric investigation on Horizontal Branch
(HB), Red Giant Branch (RGB), and Main-Sequence Turn-Off (MSTO) stars
in the Galactic globular cluster $\omega$~Centauri = NGC$\,$5139.
The central regions of the cluster were covered with a mosaic of
$F435W$, $F625W$, and $F658N-$band data collected with the Advanced 
Camera for Surveys on board the Hubble Space Telescope. The outer 
reaches were covered with a large set of $U,B,V,I-$band data collected 
with the mosaic CCD camera available at the 2.2m ESO/MPI telescope. 
The final catalogue includes $\sim 1.7$ million stars. We identified 
more than 3,200 likely HB stars, the largest sample ever collected 
in a globular cluster, and more than 12,500 stars brighter than the 
subgiant branch and fainter than the RGB bumps ($15 \le B \le 18$). 
We found that the HB morphology changes with the radial distance 
from the cluster center. 
The relative number of extreme HB stars decreases from $\sim 30$\% 
to $\sim 21$\% when moving from the center toward the outer reaches 
of the cluster, while the fraction of less hot HB stars increases 
from $\sim 62$\% to $\sim 72$\%. 
Current findings seem to support the evidence brought forward by 
Castellani et al. (2006a) that the Blue Tails, if affected by cluster 
dynamics, should be considered more a transient phenomenon rather 
than an intrinsic feature of GCs. 

We performed a detailed comparison between observed ratios of different 
stellar tracers and theoretical predictions based on evolutionary models 
constructed by adopting a canonical primordial helium (Y=0.23) content and 
metal abundances (Z=0.0002, Z=0.001) that bracket the observed spread in 
metallicity of \omc stars. We found that the empirical star counts of   
HB stars are on average larger (30\% -- 40\%) than predicted by current 
evolutionary models. Moreover, the rate of HB stars is $\sim$ 43\% 
larger than the MSTO rate. The discrepancy between the rate of HB stars 
when compared with the rate of RG and MSTO stars supports the evidence 
that we are dealing with a true excess of HB stars.  
  
Recent empirical evidence suggests the occurrence of He-enhanced stellar 
populations in \omcp. Therefore, we constructed different sets of 
evolutionary models using the same metal contents, but higher (Y=0.33, 0.42)
helium abundances. The comparison between theory and observations was 
performed by assuming a mix of stellar populations made with 70\% of 
canonical stars and 30\% of He-enhanced stars. We found that the observed 
RG/MSTO ratio agrees with the predicted lifetimes of He-mixed stellar 
populations. However, the empirical counts of HB and RG stars are once 
again systematically larger than predicted for He-mixed stellar populations. 
The discrepancy between theory and observations decreases by a factor 
of two when compared with rates predicted by canonical He content models, 
but still 15\%--25\% (Y=0.42) and 15\%--20\% (Y=0.33) higher than observed.
Furthermore, the ratio between HB and MSTO star counts are $\sim$ 24\% (Y=0.42) 
and 30\% (Y=0.33) larger than predicted lifetime ratios. This finding seems 
to be robust, since the HB/MSTO ratio presents a strong sensitivity to helium 
abundance and it is marginally affected by field star contamination.     

Finally, we briefly outline the impact that the new findings might have 
on the final evolutionary fate of low-mass stars.
\end{abstract}



\section{Introduction}\label{introduction}

The occurrence in some Galactic Globular Clusters (GGCs) of $B-$type subdwarfs 
populating the hot end of the Horizontal Branch (HB) represents a
puzzling and vigorously debated observational feature. Without doubt they are
He burning stars which have lost a large fraction of 
their H-rich envelopes. However, the mechanism governing such a peculiar 
mass-loss event is still unknown.

The discovery of similar stellar structures in the GGC \omc dates back 
to the seminal papers by Da~Costa \& Villumsen (1981) and Da~Costa, Norris 
\& Villumsen (1986). The spatial distribution of such an Extreme Horizontal 
Branch (EHB) population was investigated for the first time by 
Bailyn et al.\ (1992), who concluded that EHB stars are more centrally 
concentrated  than both normal HB stars and subgiant (SG) stars.
This, if confirmed, would represent relevant information 
about the origin of such a peculiar population. In particular, this 
would support the hypothesis of a binary mechanism ({\it e.g.\/}, either 
the merging of pairs of degenerate dwarfs, or moderately wide binaries 
experiencing mass transfer episodes just before the He flash) for 
their formation. 

However, Whitney et al.\ (1994, 1998) made use of the orbiting
Ultraviolet Imaging Telescope (UIT) facility to study \omc cluster stars 
in a far-$UV$ band ($\rm \lambda \sim 1620 \AA$), reaching the conclusion 
of a similar spatial distribution for both EHB and normal HB stars.
Finally, empirical evidence for an {\it increase\/} in the fraction of EHB 
when moving outside the cluster core (43\% vs 30\%) has been brought 
into focus by D'Cruz et al.\ (2002) discussing optical ($F555W$) and 
far-$UV$ Hubble Space Telescope (HST) data collected with three Wide Field 
Planetary Camera 2 (WFPC2) pointings. However, the same authors warn 
against the possibility of a selection effect in the $F555W-$band due 
to the crowding of the central regions. 

In spite of such long-standing interest, we still lack 
exhaustive information on the cluster HB population. As a matter of
fact, in the current literature one finds either photometry of
small cluster regions, or Color-Magnitude Diagrams (CMDs) which 
barely reach the faint  magnitudes of the EHB population. As a result, 
in the available wide-field CMDs the EHB appears as a sparse and rare 
population (see, {\it e.g.\/}, Pancino 2002; Rey et al. 2002), whereas 
D'Cruz et al.\ (2002) have already found that the EHB population 
should represent 30-40\% of the total number of cluster HB stars.

The recent detection (Anderson 2002; Bedin et al. 2004) of a well-defined 
double MS in \omc has stimulated a relevant theoretical and observational 
effort. This secondary population includes a fraction of the order of 
30\% of \omc stars and shows up as a fainter and/or bluer MS when 
compared with canonical, redder MS stars. To account for this peculiar 
evolutionary feature a substantial increase in the He abundance
has been suggested ($\Delta Y \approx 0.15$, Norris 2004; 
Piotto et al.\ 2005; Sollima et al.\ 2007). A similar large variation 
in the helium content was also suggested by Lee et al.\ (2005) to 
explain the large number of extreme horizontal branch stars present 
in \omc and in NGC2808 (D'Antona et al. 2005).  
The strong He-enrichment scenario was also supported by Maeder \&
Meynet (2006) who suggested that models of low-metallicity, massive stars
with moderate initial rotation velocities can produce stellar winds with
large He- and N-excesses. However, the self-enrichment scenario has been 
recently questioned by Bekki \& Norris (2006). By assuming that the helium 
enrichment is only caused by ejecta from stars of the red MS they found
that the suggested properties of the blue MS stars cannot be the aftermath
of a self-enrichment process from either massive AGB stars, or mass loss of
very massive young stars, or from Type II supernovae. Although the working 
hypothesis of a mixed He-enriched population accounts for some morphological 
empirical evidence, we still lack a detailed comparison between observed star 
counts and predicted lifetimes based on both canonical and He-enhanced 
evolutionary tracks and horizontal branch models.     
 
In this paper we investigate the radial distribution of HB, RG, and
main-sequence stars. In order to investigate star counts over a 
substantial fraction of \omc we use two different data sets. 
Multiband ($F435W$, $F625W$, $F658N$) data collected with Advanced Camera 
for Surveys (ACS) on board the HST are adopted for the central cluster 
regions, while multiband 
($U,B,V,I$) ground-based data collected with a mosaic CCD camera 
(WFI@2.2m ESO/MPI) have been adopted for the external cluster regions.      
In \S 2, we discuss the two different photometric data sets, together 
with the strategy we adopted to perform the photometry. The approach we
employed to transform stellar positions from pixels to equatorial 
coordinates and the cross-identification of the optical catalogue with 
the near-infrared and the proper motion catalogues are presented in 
\S 3. Empirical estimates of star counts for HB and RG stars are 
presented in \S 4 together with the comparison with evolutionary 
prescriptions that cover a broad range of metal abundances and 
plausible changes in cluster ages. \S 5 deals with the comparison 
of observed HB, RG, and MS star counts with their evolutionary 
lifetimes. The impact that a significant increase in the helium 
content has on predicted lifetimes of quoted evolutionary phases 
is discussed in detail in \S 6. A brief summary of the results and 
a few viable developments to further constrain the evolutionary 
properties of stellar populations in \omc are outlined in \S 7.

\section{Observations and data reduction}

Photometric data discussed in this investigation belong to two different 
sets from both space (HST) and ground based telescopes. 
Multiband ($U,B,V,I$) data have been collected with the mosaic ($2\times 4$) 
CCD camera Wide Field Imager (WFI, Baade et al. 1999) available at the 
2.2m ESO/MPI telescope (ESO, La Silla). 
These data have been retrieved from the ESO science archive and include 
8 $U$, 39 $B$, 51 $V$, 26 $I$ pointings. The data include both shallow 
and relatively deep images, with exposure times ranging from 1 to 
300 seconds for the $B$, $V$, and $I$ bands, and from 300 to 2,400 seconds 
for the $U$ band. The total exposure time per band for the different 
pointings is 7500 s ($U$), 3431 s ($B$), 3969 s ($V$), and 2156 s ($I$). 
The data were collected in several observing 
runs ranging from 1999 to 2003. During this period two filters were 
changed: data secured before 2002 were collected with the 
ESO filters $U/38_{ESO841}$ and $B/99_{ESO842}$, while later ones with the 
ESO filters $U/50_{ESO877}$ and $B/123_{ESO878}$. These data were 
obtained in good seeing conditions, and indeed the mean seeing ranges 
from $\sim$ 0\farcs60  for the $I-$band to $\sim$ 1\farcs1 for the 
$U-$band data.  Table 1 gives, from left to right, the frame identification, 
the coordinates, the modified Julian date, the exposure time, the filter, 
and the seeing for each WFI pointing.   
The field of view (FOV) covered by each pointing is $34\times 33$ arcminutes 
squared, and the FOV covered by the entire dataset is $42\times 48$ arcminutes
across the center of the cluster. The large solid black polygon in Fig. 1 
shows the cluster area covered by this data set. The raw frames were 
pre-reduced by using standard IRAF procedures. 

Multiband ($F435W$, $F625W$, $F658N$) photometric data were collected with 
the ACS on board HST, and retrieved from the 
HST archive. These data include nine pointings across the center of the 
cluster (see blue solid polygons in Fig. 1). The $3\times 3$ mosaic covers a
field of view of $\approx 10\arcmin \times 9.5\arcmin$. Four images per
field were acquired in three different bands. Data in the $F435W$ and 
in the $F625W$ band were secured with one shallow ($8$s) and three 
deep ($340$s each) exposures, while the exposure time for the four 
$H_\alpha$ ($F658N$) images was 440s each. The raw frames were 
pre-reduced by using the standard HST pipeline.  
By assuming a King (1962) profile and the structural parameters (core 
radius, $r_c=2.6$ arcmin; tidal radius, $r_t=45 $ arcmin) for 
\omcp\footnote{No 
general consensus has been reached yet on the structural parameters 
of \omc, the GGC catalogue by Harris (1996) gives $r_c=1.4$ and 
$r_t=57$ arcmin, while Meylan et al. (1995) give $r_c=3$ and $r_t=51$ arcmin.}  
given by Trager, King, \& Djorgovski (1995), we estimate that the current 
photometry accounts for approximately the 99\% of the light of the cluster.

To improve the photometric accuracy, carefully chosen selection criteria were 
applied to pinpoint a large number ($\approx$200) of isolated point-spread function 
stars across the individual chips/frames, and several different reduction strategies 
were used to perform the photometry ({\tt DAOPHOT/ALLSTAR}). The observed ground-based 
($124\times 8$) and space (108) photometric catalogues were rescaled to a common 
geometrical system with {\tt DAOMATCH/DAOMASTER}. The entire mosaic of HST data was 
simultaneously reduced with {\tt DAOPHOTII/ALLFRAME} and the final catalogue 
includes $\sim 1.32\times10^6$ stars with at least one measurement in two 
different photometric bands. The photometry was kept in the Vega system 
following the prescriptions suggested by Sirianni et al.\ (2006, and references 
therein).
The main difference with the investigations by Freyhammer et al. (2005) and 
Monelli et al.\ (2005), which are based on the same data, is that they performed 
the photometry on individual pointings and adopted slightly different zero-points 
in the absolute calibration.  

The ground-based data were divided into shallow (89) and deep (35) pointings and 
each set was simultaneously reduced with {\tt DAOPHOTII/ALLFRAME}. The final 
catalogue includes $\sim 6.4\times10^5$ stars with at least one measurement in 
two different photometric bands. The relative 
and absolute photometric calibration of ground-based instrumental magnitudes was 
performed following the strategy suggested by Monelli et al. (2003) and Corsi 
et al. (2003). To accomplish this non-trivial objective we used a large set 
($\sim 30,000$ stars) of new multiband ($U$,$B$,$V$,$I$) local photometric standard 
stars (Stetson 2000\footnote{See also http://cadcwww.hia.nrc.ca/cadcbin/wdb/astrocat/stetson/query/}; 
Stetson et al. 2006, in preparation). The sky area covered by the 
standard stars is approximately of $33\times 39$ arcmin squared and includes 
a substantial fraction of the current scientific data. 
The comparison between the current calibrated data and the local standard stars 
observed by Walker (1994) indicates that the agreement in the absolute 
zero-points is of the order of 0.03 mag ($B-$band) or better ($V,I-$band).  

The entire photometric catalogue includes $\sim 1.7\times10^6$ stars and it is, 
to our knowledge, the largest multiband data set ever collected for a Galactic 
Globular Cluster (GGC). Data reduction required several months of CPU time on 
a dedicated 64-bit server. Fig. 2 shows a color image of the central cluster 
regions with almost the same spatial resolution as the original ACS data. 
The color of individual stars is based on $U,B,V,I-$band data collected 
with the WFI.   
Note that the crowding of the innermost regions is a major problem also for a
globular cluster with a modest central density
($\log \rho = 3.12\;L_\odot \, pc^{-3}$, Harris 1996).

In order to provide a composite CMD including both ground-based and space 
data, we transformed the $F435W$ 
magnitudes into the standard $B-$band. By selecting common stars with 
$14.5 \le B \le 22.5$ we found the following calibration 
$B = F435W + 0.03(\pm 0.02) - 0.0015(\pm 0.001) F435W$. Furthermore, 
we transformed ground-based $V-$ and $I-$band data into $F625W$ magnitudes 
as $F625W = V\times0.544 + I\times0.455$. The two sets of magnitudes agree quite well,  
and indeed we found that the difference is within an rms of $0.03$ mag. 
For the stars with measurements both in the ACS and in the WFI dataset
we adopted the former one.
Figure 3 shows the CMD in $B$, $B-F625W$. 
Stars plotted in this figure have been selected according to photometric 
accuracy ($\sigma_B,\sigma_V,\sigma_{F625W},\sigma_I < 0.03$) and the ``separation 
index'' ({\tt sep > 3})\footnote{The `separation index' quantifies the degree
of crowding (Stetson et al. 2003). The current $\tt sep$ value 
corresponds to stars that have required a correction of less than 6\% for light
contributed by known neighbours.}. A good fraction of the bright Red Giant
Branch (RGB, $B\sim 15.5$, $B-F625W \ge 1.4$) stars mainly comes from the WFI
photometry, while faint main sequence stars come from the ACS photometry. A
glance at the data plotted in this figure shows that relatively fast
evolutionary phases such as the Asymptotic Giant Branch (AGB, $B\sim 14.5$,
$B-F625W\sim 1.4$), the Blue Straggler sequence (BS, $16.5 \le B \le 18.5$,
$0.3 \le B-F625W \le 0.8$), and the bright region of the White Dwarf (WD)
cooling sequence ($B > 21$, $B-F625W\sim -0.2$) are all sampled well.  
However, we can not exclude that we are missing bright AGB stars
($B\le 13$, $B-F625W \ge 2.4$)  and a fraction of RG stars located close to the
tip of the RGB ($B\sim 13-13.5$, $B-F625W \sim 2.3-2.4$).

The Horizontal Branch (HB), in particular, appears very well populated and 
shows a morphology quite similar to the blue HB of NGC2808 (Bedin et al 2000; 
Castellani et al. 2006a). Following the scheme adopted for NGC2808 stars, the blue 
HB tail can be divided into three different subgroups. Stars hotter than the RR Lyrae
instability strip and brighter than the first gap ($B \le 16.5$, EBT1),   
stars located between the first and the second gap ($16.5 \le B \le 17.8$, EBT2), 
and stars fainter than the second gap ($B > 17.8$, EBT3).  
The current limits (see solid lines in Fig. 7) have been fixed 
arbitrarily, since the first gap---unlike the case of NGC2808---is not 
well defined in the optical bands.  

\section{Cluster members: cleaning the herd}

In order to investigate the HB morphology as a function of the radial 
distance we devised the following approach:  

{\em i)} The optical catalogue was transformed into the equatorial system by
using IRAF's IMMATCH package to establish the spatial transformation using
a subsample of 704 stars in common with the van Leeuwen et al.\ (2000)
catalogue of proper motions and membership probabilities in \omcp.
The latter catalogue covers a field measuring 47\arcmin\ by 45\arcmin\ on the
sky, which was offset 9\farcm3 west and 3\farcm2 south of the cluster
center (center ICRF position: $\alpha=201\fdg69065$, $\delta=-47\fdg47855$)
to include a part of the halo. The mutual overlap with our field is 
approximately 36\farcm9 by 42\farcm9. The limiting magnitude of the 
9847 stars in this
catalogue is $V=16.0-16.5$, and the
precision of the proper motions is 0.1--0.65 mas yr$^{-1}$, 
depending on stellar brightness. The r.m.s.\ of our transformation is 
0\farcs09 and 0\farcs05 in $\alpha$ and  $\delta$, respectively. 
We rejected matches for objects with separations exceeding 1\farcs0 and 
identified $\sim 6,300$ stars in common. Among them, 280 stars (open 
circles in Fig. 5) have a probability smaller than 10\% of being 
cluster members (van Leeuwen et al. 2000).

{\em ii)} Subsequently, the transformed equatorial coordinates of our
optical catalogue were matched with those in the near-infrared (NIR) Two
Micron All Sky Survey (2MASS, Cutri et al. 2003)
catalogue\footnote{See also http://www.ipac.caltech.edu/2mass/releases/allsky/}.
We rejected matches for objects with separations exceeding 0\farcs4 and
identified $\sim 15,000$ stars in common. For these objects we have at least
three optical and three NIR ($J,H,K$) magnitudes. In order to split field
and cluster stars we used the $U-J$ vs $B-H$ color-color plane (see 
top panel of Fig. 4) together 
with several optical-NIR CMDs, namely $B, U-J$, $V, B-H$, and $I, U-I$. 
We did not use the $K$-band photometry because it is less accurate in 
the faint magnitude limit ($B \ge 17$). 
On the basis of these CMDs we provided a preliminary cut of the region of this plane
where \omc RG stars are located by using the stars distributed in an annulus close
to the cluster center ($2.5 \le r \le 10$ arcmin). We performed a linear fit of
the $U-J$ color as a function of the $B-H$ color for the selected stars and
then we estimated for each star in the sample the distance in color from
the fitting line. We divided the sample in different $B-H$ color
intervals and for each bin we fit the difference in $U-J$ color with a
gaussian. The peaks and the sigmas of the different gaussians were adopted to
improve the location of the fitting line and the selection of likely \omc stars.
The procedure was repeated a few times by slightly changing the cuts in $B-H$ color.
Data plotted in the middle and bottom panel of Fig. 4 show the last gaussian
fits we performed, while the blue line in the top panel shows the final fitting
line. The likely cluster RG stars (black dots) appear to be distributed along
a well-defined sequence, while field objects form either two cooler plumes
(red dots) or a hotter tail (green dots) for $1.5 \le U-J \le 4$.
Likely cluster HB stars (blue dots) are located in the left bottom corner 
due to the strong sensitivity of these colors to the effective temperature. 
The $U$ magnitudes also show some dependence on surface gravity, since the
Balmer convergence and continuum are included in that bandpass: high-luminosity
stars are fainter in $U$ than low-luminosity stars of the same long-wavelength
colors.
A few dozen field objects present the same colors as HB stars, but they 
are significantly fainter/brighter than HB stars. Therefore we also selected 
HB stars on the basis of a magnitude criterion ($14.5 \le B \le 16.1$).
Note that to avoid systematic drifts in the selection of the different stellar 
samples we required that likely cluster RG and HB stars be 
located inside the expected magnitude and color intervals in the different 
optical-NIR CMDs.

The left panel of Fig.~5 shows the $B, U-H$ CMD of the cluster 
sample for candidate members, while the right one shows the CMD of 
the field objects ($\sim 1600$). 
Although the criteria to select the different samples rely on photometric
properties and partially on proper motions, data plotted in the left panel
indicates that field objects have been properly subtracted.
The AGB stars display a well-defined split from the 
RG stars, and the bright RG stars of the {\em anomalous branch} ($\omega 3$, 
Rey et al. 2000; Pancino et al. 2000; Freyhammer et al. 2005) can be 
easily identified (see the arrows).

Furthermore, data plotted in the right panel show the typical 
color distribution of field stars (see, {\it e.g.\/}, Gilmore et al. 1990). 
However, the small over-density of objects located 
at $B\sim 15$ and $U-H\sim1.5$ suggests that the current selection 
includes a few cool cluster HB stars in the sample of field 
objects. As an independent test of the appropriateness
of the current selection criteria, Fig. 6 shows the radial distribution 
of cluster stars (solid line) and field objects (dashed line). 
We accounted for the fact that \omc is ellipsoidal with a position 
angle of $\approx 100\deg$ (van de Ven et al. 2006, and 
references therein) and the positions of individual objects 
were estimated as the distances projected along the major 
and minor axis, respectively. 
Data plotted in this figure show that the ``field'' objects show 
a mild increase when approaching the cluster center. However, this 
field sample is approximately 10\% as large as the sample of cluster 
stars. Note that the decrease in the number of stars close to the 
cluster centre is due to the limited spatial resolution of the 
2MASS catalogue. Moreover, galaxy counts based on multiband 
photometric data collected by the Sloan Digital Sky Survey 
(SDSS) suggest that the expected number of field galaxies per square 
degree with $B \le 18$ is negligible ($\approx 30$, 
Yasuda et al. 2001)\footnote{Note that to perform this estimate we accounted 
for the difference in magnitude between the Johnson-Cousins system and the 
SDSS system (see, {\it e.g.\/}, Jordi, Grebel, \& Ammon 2006).}. 
This evidence indicates that the conclusions of this investigation are 
at most marginally affected by the criteria adopted to distinguish field 
objects from cluster stars. For a more detailed discussion on the approach 
adopted to split field and cluster stars see Calamida et al. (2007, in 
preparation).  

Once we removed the field objects from the optical-NIR 
catalogue we estimated the total flux in the $B-$band from the stars 
fainter than $B\sim 13$ and brighter than $B\sim 22$. Note that to 
estimate the total flux we did not apply selection criteria to the 
photometric catalogue, apart from the rejection of probable field objects.  
Moreover, in this magnitude range the completeness of the two photometric
catalogues as a function of the radial distance is very similar.
Then we divided the photometric catalogue into three concentric 
regions in such a way that each of them includes one third of 
the flux of the entire sample. We found that the three regions are given by
$r \le r_\alpha =204.3$, $r_\alpha < r \le r_\beta =458.6$, 
and $r > r_\beta$ arcsec. Note that the current radial cuts marginally 
depend on the limits in magnitude adopted to estimate the fractional flux. 
An increase/decrease of one magnitude in the bright cut causes a change 
of $\pm 3$ arcsec in $r_\alpha$ and of $\pm 11$ arcsec in $r_\beta$. 
A similar change in the faint cut affects the values of $r_\alpha$ 
and $r_\beta$ by approximately one arcsec. For the stars located in the 
innermost regions we used only the HST photometry, while for the middle 
region we adopted either HST photometry ($r \le r_{acs}\sim 400$ arcsec) 
or ground-based photometry ($r_{acs} < r \le r_\beta$). 
Fig. 7 shows, from left to right, the CMDs of space ($B$, $B-F625W$) and 
ground-based ($B$, $B-V$) data for annuli at increasing radial distance.

Data plotted in this figure show that the EBT2 region appears 
less populated than EBT1 and EBT3 regions in all three radial 
zones.  Moreover, stars in the EBT3 region appear to  
decrease more rapidly toward the outskirts of the cluster when compared 
with stars in the EBT1 and in the EBT2 region. To express this circumstantial 
evidence on a more quantitative basis, Table 2 lists the star counts of the 
different sub-samples as a function of the radial distance. Note that the 
uncertainties given for the relative fractions of individual sub-samples only 
account for the statistical errors in the star counts. 
{\em The stars plotted as black dots in Fig. 7 have been selected on the basis
of the separation index, photometric errors, and sharpness. On the other hand,
we did not apply any selection criteria to count the different stellar
sub-samples (RG, red dots; HB, blue dots; $\omega_3$-branch, green dots;
MS, cyan dots) apart from the upper and lower magnitude limits, and the
rejection of field objects.}
We ended up with a sample of 3311 likely HB stars, which is by far 
the largest sample of HB stars measured in a GGC to date. On the whole, one 
finds that the relative number of HB stars (as normalized to the total 
$B$-band flux) steadily decreases when moving 
from the center to the outermost regions. This trend applies to individual 
sub-samples, therefore it can hardly be explained as an observational bias.

Moreover, the ratio between stars in the EBT3 and the total number of HB 
stars is roughly equal, within the errors, for $r \le r_\alpha$ 
($N(EBT3)/N(HB)\sim 0.28\pm 0.02$) and for $r_\alpha < r \le r_\beta$ 
($N(EBT3)/N(HB)\sim 0.32\pm 0.02$) and decreases to $\sim 0.21\pm 0.02$ 
in the outermost region, while the fraction of stars in the EBT2 region 
is constant across the cluster ($N(EBT2)/N(HB)\sim 0.08\pm 0.01$). 
On the other hand, the fraction of stars in the EBT1 region is
$N(EBT1)/N(HB)\sim 0.62\pm 0.03$ for $r \le r_\beta$ and increase
to $N(EBT1)/N(HB)\sim 0.72\pm 0.04$ for $r > r_\beta$. 
The anonymous referee noted that the difference in the radial distribution
between stars in the EBT1 and in the EBT3 region is further strengthened by
the evidence that the ratio $N(EBT3)/N(EBT1)$ for $r \le r_\alpha$ differs
at $5.9\sigma$ level ($T^2$ statistics) with the same ratio for $r > r_\beta$.
A glance at the $B-$band luminosity functions (LFs) 
of HB stars plotted in Fig.~8 as a function of the radial distance further 
confirms this radial trend. Note that the current space- and ground-based 
photometry is significantly deeper than the selected HB stars. Therefore 
current star counts should be minimally affected by completeness problems. 
Moreover, crowding effects would act in the opposite direction, since they should 
spuriously reduce the counts of EBT3 stars more in the region closer to the center 
than in the outskirts of the cluster.

However, to provide firm constraints on deceptive systematic errors that 
might affect the current HB star counts (crowding, saturated stars, limiting 
magnitudes) we performed the same selections by using different colors 
for both space ($B$, $B-F658N$) and ground-based ($B$, $U-V$) photometry. 
The new star counts are listed in Table 2 and they agree quite well---within 
the errors---with those quoted above. Moreover, HB stars plotted in Fig. 9 
display the same trend concerning the radial gradient of the three EBT 
regions. Current findings support the results obtained by 
Bailyn et al. (1992) concerning the steady decrease in the number of 
EHB stars when moving toward the outer reaches of the cluster. They also 
support the results by D'Cruz et al. (2002) concerning the relative 
fraction of EBT3 stars ($\sim 27\pm1$\% vs 30\%--40\%).

\section{HB versus Red Giant Branch stars}

A relevant question concerning the occurrence of HB ``blue tails" in 
globular clusters is to assess whether observed star counts along the 
HB agree with the normal evolutionary sequence in which HB stars are 
the predicted progeny of RG stars that have just experienced 
the He-core flash at the tip of the RGB. There are indeed evolutionary 
scenarios suggesting that either a fraction of HB stars approaches the 
blue tail along a different evolutionary path, or a fraction of RG 
stars misses the HB evolutionary phase altogether due to a virtually 
complete removal of the envelope.  

It has already been suggested that the hot EBT3 group could be the 
aftermath of a quite different evolutionary path, {\it e.g.\/}, the coalescence 
of two low-mass He-core WDs (Iben 1991, and references therein). In such 
a case the stars in the EBT3 group will appear as an addition to the 
normal and legitimate HB, and in turn, the total number of HB stars 
should exceed the number expected from the ongoing transition from RG 
into HB structures.

However, according to an alternative suggestion  advanced by
Castellani \& Castellani (1993), but see also D'Cruz et al.\ (1996), 
Brown et al. (2001), and Cassisi et al.\ (2003), the stars in the 
EBT3 region might be the result of an episode of extreme mass 
loss---possibly caused by stellar encounters or compact binaries 
(Ivanova et al. 2006)---before 
the onset of the He-core flash. This phenomenon 
can drive the stellar mass very near to the lower mass limit for central 
He ignition. The RG stars {\it above\/} this limit undergo a late He-flash 
either during the contraction toward their He-WD structure or in the approach 
to their WD cooling sequence.  These stellar structures have been christened 
in the literature as {\em hot He-flashers} (Castellani \& Castellani 1993; 
Sweigart 1997) and they have been photometrically (Momany et al.  2004) and
spectroscopically (Moehler et al. 2002; Moehler \& Sweigart 2006, and references 
therein) identified in several GCs.  
However, it would not be a surprise if the same significant 
mass-loss mechanism also drives a fraction of the RG stars {\it below\/} 
the limit for central He ignition. This would imply that a fraction of 
RG stars will miss the He-core flash, thus directly evolving into cooling 
He-core WDs rather than spending time as HB stars (Castellani, Castellani, 
\& Prada Moroni 2006). 
The key consequence of such an evolutionary path is that one should 
observe a lower number of RG stars than predicted by evolutionary models. 
Moreover, according to the fraction of RG stars that after the heavy mass-loss 
episode are either {\it above\/} or {\it below\/} the limit for central He ignition 
one should observe either an increase in the number of extreme HB stars or in 
the number of stars in the bright region of the WD cooling sequence, relative 
to the total number of RG stars. It goes without saying that RG 
stars with stellar masses uniformly distributed across this limit should 
produce an increase in the number of both EHB stars {\em and\/} bright WDs
(Castellani et al. 2006b).

One may explore such a problem by recalling that at advanced
stages of evolution the number of  stars observed in a given
evolutionary phase is expected to be proportional to the time
spent by stellar models in that given phase. We will check the
number of HB stars measured in \omc by using the RG stars as a 
``reference clock''. We counted RG stars in the magnitude range 
$15.0 \le B \le 17.8$. We selected this magnitude interval 
to secure a sizable sample of cluster stars in an evolutionary 
phase whose lifetime is minimally affected by both chemical 
discontinuity (RGB bump) and mass-loss efficiency (tip of the RGB). 
Star counts listed in Table 3 show that RG stars present the same 
radial trend as HB stars, {\it i.e.\/}, their numbers 
steadily decrease (relative to the total $B$-band flux) 
when moving from the center to the outermost regions. Owing to 
the photometric precision and to the large sample of measured 
stars, data plotted in Figs.~7 and 9  show a well-defined 
$\omega 3-$branch in the three different annuli (green dots). 
Star counts listed in Table 3 also support the finding by 
Pancino et al.\ (2000) that this stellar population is 
approximately 3--4\% of the entire population. There 
is weak evidence that this stellar population might increase 
when moving toward the outskirts. But on the whole, RG stars on the 
$\omega 3-$branch are less than 5\% of the entire RG population 
and they will be neglected. 

In order to provide a more quantitative analysis concerning 
star counts along the RGB, we counted RG stars in bins of 
one $B-$mag. This count was performed for eleven bins by 
adopting a shift of 0.1 in the magnitude range $16.6 \le B \le 17.6$
(see columns 3 to 5 in Table 4). Fig. 10 shows the RG star counts 
based on $B$, $B-F625W$ (ACS@HST) and $B$, $B-V$ (WFI@2.2m) CMDs as 
a function of the $B-$mag. Interestingly enough, RG stars do not 
show any significant radial trend for $r \le r_\beta$, and indeed 
the star counts attain very similar values over the entire magnitude 
range.  On the other hand, star counts in the external annulus are 
systematically smaller than the inner ones in all the bins. 
The same outcome applies if we use the RG stars selected in the 
CMDs based on $B$, $B-F658N$ (ACS@HST) and $B$, $U-V$ (WFI@2.2m) 
CMDs (dashed lines). 
The selection of the adopted radial annuli is based on the total $B-$band flux
emitted by the different stellar populations in \omcp. In order to validate
the current approach, we compared the observed density profile of the entire
stellar population with a standard isotropic single-mass King model with
$r_c=155$ arcsec and $c=\log r_t/r_c=1.24$ (Trager et al. 1995; Ferraro et al. 2006)
and we found a very good agreement. Therefore, the different populations should
contribute to the total flux with the same number of stars. The difference we found
in the outermost annulus might be caused either by the RG stars brighter than the HB
or by the AGB stars. The star counts of these evolutionary phases will be discussed
in a forthcoming paper.

In order to establish whether the number of RG stars is peculiar 
either in the two inner annuli or in the external one we need to 
compare empirical star counts with predicted evolutionary times.  
Inspection of recent evolutionary prescriptions  
(Pietrinferni et al.\ 2006), shows that in the lower portion 
of the RG branch ({\it i.e.\/}, below the luminosity of the RGB bump and 
above the subgiant branch), this time is largely independent of 
cluster metallicity and of cluster age for ages ranging 
from 9 to 13 Gyr. The same outcome is also supported by empirical 
evidence, and indeed Stetson (1991) showed that cluster LFs, once 
shifted so that the Turn-Off magnitudes are coincident, marginally
depend along the RGB on iron and $\alpha$-element abundances as 
well as on cluster age and initial mass function.     

In order to provide more quantitative estimates 
of this dependence we constructed several evolutionary tracks by 
adopting different stellar masses and chemical compositions. 
Current evolutionary tracks\footnote{Several sets of evolutionary 
tracks and isochrones are available at the URL 
http://astro.df.unipi.it/SAA/PEL/Z0.html} have been computed with an
updated version of the FRANEC evolutionary code (see, {\it e.g.\/}, Chieffi \&
Straniero 1989) and by adopting the radiative opacities made available 
by the Livermore group (Iglesias \& Rogers 1996) and updated nuclear 
cross sections (Cariulo, Degl'Innocenti, \& Castellani 2004). 
Current evolutionary tracks account for element diffusion (Ciacio, 
Degl'Innocenti, \& Ricci 1997) with diffusion coefficients from 
Thoul, Bahcall \& Loeb (1994), and neglect mass loss during 
hydrogen- and helium-burning phases. Convective transport in the envelope 
regions is treated with the mixing-length formalism in which the 
parameter $\alpha$ defines the ratio between the mixing length 
and the local pressure scale height. We adopted $\alpha$=2.0, which 
reproduces---with the currently 
adopted color transformations---the observed RG branch colors of 
the Galactic globular clusters (Cariulo et al. 2004). Stellar 
structures during He-burning phases have been constructed according 
to the canonical prescriptions for ``induced semiconvection'' 
(see, {\it e.g.\/}, Castellani et al.\ 1985).

To bracket the chemical compositions of the bulk of the stellar populations in
\omc (Norris, Freeman \& Mighell 1996; Suntzeff \& Kraft 1996; Hilker et al.\ 2004;
Sollima et al.\ 2005; Piotto et al.\ 2005; Sollima et al. 2006) we adopted
Z=0.0002 and Z=0.001.
Although precise empirical estimates of the metallicity distribution in 
\omc are available, we decided to use only two metal abundances, since the 
evolutionary lifetimes we plan to use display only a mild dependence on this 
intrinsic parameter. Moreover, to 
account for the age effect we adopted two stellar masses $M/M_\odot=0.8$ 
and  $0.85$. These stellar masses imply a difference of $\approx 2\,$Gyr 
in the turnoff age. Data plotted in Fig.~11 and evolutionary times listed in 
Table 5 show that an increase in metal abundance of a factor of five 
and a decrease in cluster age of $\approx 20$\% cause on average a 
change in the RG lifetime of the order of 10\%. This evidence strengthens 
the use of RG stars as a reference clock in stellar systems showing a 
spread in metal abundance and/or in stellar age such as $\omega$~Cen.
Current evolutionary models for Z$\leq$0.001 have been constructed by 
adopting an original helium abundance of $Y\sim 0.23$. By taking 
into account a conservative estimate for the uncertainty of this input 
parameter ($0.23 < Y < 0.25$; Salaris et al. 2004, Spergel et al. 2006) 
we obtain a variation of the RGB lifetime, 
for a stellar mass of $\sim 0.8 M_{\odot}$, less than 10\%. Similarly, 
the error in the RGB lifetime due to the uncertainties in the 
input physics is of the order of $\sim 10$\% (Cassisi et al. 1998).
On the other hand, the HB lifetime is marginally affected by a variation 
in the original helium abundance within plausible ranges (see, {\it e.g.\/},
Bono et al.\ 1995). Cassisi et al.\ (2001) estimated a variation of 
$\sim  10$\% of the HB lifetime due to the uncertainty in the adopted 
input physics.

Theoretical predictions have been transformed into the observational 
plane by adopting the bolometric corrections and the color-temperature 
transformations provided by Bessell, Castelli \& Plez (1998). In particular, 
we adopted atmospheric models with an $\alpha-$enhanced composition 
([$\alpha$/Fe]=0.4)\footnote{The entire set of atmosphere models is available 
at the URL {\tt http://wwwuser.oat.ts.astro.it/castelli/colors/bcp.html}.}.   
These predictions are only available for surface gravities 
$\log g \le 5$. In order to transform extreme HB models into the 
observational plane we linearly extrapolated by 0.5 dex the surface 
gravities out of the available range.
The evolutionary times along the RGB have been estimated by adopting 
the same approach adopted for empirical star counts together with a 
distance modulus of $\mu_0 = 13.70\pm0.06\pm0.06$ (Del Principe et al. 2006),  
and a reddening of $E(B-V)=0.11\pm 0.02$ (Kaluzny et al. 2002; 
Lub 2002; Calamida et al. 2005). By adopting the reddening law 
from Cardelli et al. (1989) we estimate that the apparent distance 
modulus in the $B$ band is $(m-M)_B = 14.16\pm 0.06 \pm 0.09$ mag.  
The systematic error accounts for the uncertainty both in the absolute 
zero point of the $B-$band and in the cluster reddening.    

Fig.~12 shows the comparison between star counts and theoretical 
predictions at fixed stellar mass (top) and chemical composition 
(bottom). Evolutionary lifetimes have been normalized to observations 
in the magnitude bin $m_B=16.9-15.9$ ($M_B=2.74-1.75$). We selected 
this bin to avoid the possible interference of the RGB Bump. This 
intrinsic feature is indeed systematically brighter than our adopted
magnitude limit over a broad range of stellar metallicities (see, 
{\it e.g.\/}, Table 1 in Caputo \& Cassisi 2002). Theoretical
predictions present a marginal dependence on metal abundance,  
and indeed predicted RG lifetimes for Z=0.0002 (red lines) and for Z=0.001 
(blue line) attain quite similar values. Note that to avoid a complete 
overlap the lifetimes for Z=0.001 have only been compared with the 
star counts of the outermost annulus. The same outcome applies to the 
dependence on the cluster age, and indeed a decrease of 2 Gyr 
($M=0.80 M_\odot$, red lines vs $M=0.85 M_\odot$, green line) cause a 
marginal difference in the predicted RG lifetime. Once again to avoid 
the complete overlap the predictions for $M=0.85 M_\odot$ have only 
been compared with the star counts of the outermost annulus.

Data plotted in Fig.~12 indicate that theory and observations agree 
quite well in the two external annuli, while counts of the faintest giants attain 
slightly larger values in the inner annulus. This discrepancy can be caused 
neither by an increase in metal abundance of a factor of five (top, 
blue line) nor by a decrease in stellar age of the order of 20\%  
(bottom, green line).   

Simple physical arguments suggest that the ratio between the number of 
HB stars and the number of RG stars should scale according to the ratio 
of their relative evolutionary lifetimes. Data plotted in Fig. 11 
indicate that HB lifetimes for extremely hot HB stars (EBT3, 
$T_e\sim 29,600$ K, $M/M_\odot=0.518$) are of the order of 
94 Myr for Z=0.0002 and of 101 Myr for Z=0.001 ($T_e\sim 28,500$ K, 
$M/M_\odot=0.510$). This evolutionary lifetime for hot HB stars (EBT2) 
decreases to 84 Myr (Z=0.0002, $T_e\sim 18,200$ K, $M/M_\odot=0.570$) 
and to 90 Myr (Z=0.001, $T_e\sim 18,200$ K, $M/M_\odot=0.551$), 
while for less hot HB stars (EBT1) the evolutionary lifetime 
decreases to 76 Myr (Z=0.0002, $T_e\sim 10,500$ K, $M/M_\odot=0.670$) 
and to 82 Myr (Z=0.001, $T_e\sim 10,000$ K, $M/M_\odot=0.620$), 
respectively.  Thus, ``typical'' average lifetimes are $\sim$98, 87,
and 79~Myr for EBT3, EBT2, and EBT1, respectively. Note that the three 
effective temperatures, and in turn the lifetimes, were selected to be 
representative of the three EBT regions (see Fig. 7) and of the peaks 
in the HB luminosity function (see Fig. 8).

In order to account for such an occurrence we estimated the arrival 
rate of stars onto the HB in the three EBT groups, namely
$r(HB) = N(EBT3)/98 + N(EBT2)/87 + N(EBT1)/79$. Note that to account for 
the possible spread in metal abundance the lifetime typical of the EBT3 
group was estimated as the average $t_{EBT3}$ lifetime for Z=0.0002 and 
Z=0.001 stellar structures. The same approach was adopted for the other 
HB groups and for the lifetime of RG structures. The use of $r(HB)$ represents 
the key advantage for overcoming the problem of the change in the predicted 
lifetime according to the HB morphology. Thus, the arrival rate of HB stars provides 
the opportunity to use, in the comparison between theory and observations, 
specific evolutionary time scales. The rates of HB stars in the three annuli 
are listed in column (6) of Table 2. As a whole, we found that the total rate 
of HB stars in \omc is $\sim 39$ stars per Myr. A similar rate was also estimated 
for RG stars by using the RG counts in the magnitude bins $16.60 \le B \le 17.60$ 
(see Table 4) and the evolutionary lifetimes listed in columns (3) and (4) 
of Table 5. Figure 13 shows the ratio between the HB and the RG rates.
Interestingly enough, we found that the empirical HB star counts are on 
average between 30\% and 40\% more numerous than predicted by current    
evolutionary models. 
The mean in the three different radial annuli are $1.416\pm0.047$,
$1.336\pm0.048$, and $1.339\pm 0.030$ in the $B,B-F625W$ (ACS) and
$B,B-V$ (WFI) CMDs, while they are $1.373\pm 0.043$, $1.290\pm 0.046$, and
$1.385\pm 0.027$ in the $B,B-F658N$ and $B,U-V$ CMDs. The current estimates,
taken at face value, present the mild evidence that the ratio between HB
and RG stars appears, on average in the different CMDs, larger in the innermost 
annulus ($\sim 1 \sigma$ level). However, the current ratios do not allow us to
determine whether such a discrepancy is caused by a decrease in the rate
of relatively faint RG stars or by an increase in the rate of HB stars.
 
\section{HB and RGB versus Main Sequence stars}

The ratio between the rates of HB and RGB stars attains very plausible values as a 
function of the radial distance. However, to constrain on a quantitative basis 
whether the star counts of HB and RG stars present any peculiar radial trend 
we need to use MS stars as an independent clock. In particular, we selected  
MS stars just below the turnoff (TO) region in the magnitude bin 
$18.65 \le B \le 19.15$ mag. The reason why we selected this region is 
twofold: {\em i)} empirical and theoretical evidence suggests that it 
marginally depends on the initial mass function (Zoccali \& Piotto 2000);
{\em ii)} the magnitude range is quite similar to the magnitude of  
EHB stars, hence the two samples also present a very similar 
completeness; {\em ii)} stellar structures in this magnitude range 
present a small spread in mass, and indeed $\Delta M \le 0.025 M_\odot$.
Fig.~14 shows the ratio 
between RG stars and MS stars in the three different annuli. The different 
ratios attain quite similar values over the entire magnitude range, thus 
suggesting homologous radial distributions for RG and MS stars. 
Interestingly enough, the predicted ratios for $M=0.80 M_\odot$ 
and different metal abundances (red and blue lines) between MS 
(see open and filled circles in Fig. 11) 
and RG (see data listed in columns 3 and 4 of Table 5) lifetimes are in 
reasonable agreement with empirical ratios. Note that the theoretical 
ratios were not normalized to observed ratios and that no vertical shift 
was applied. The comparison between theory and observations indicates a 
mild excess in the number of observed RG stars. If we assume as a typical 
predicted ratio the mean of the two different metal abundances we found 
that the discrepancy between theory and observations on average ranges 
from $\sim 10$\% to $\sim 15$\% when moving from brighter to fainter 
RG stars. The same outcome applies if we use the predicted ratios based on 
younger stellar structures, {\it i.e.\/}, lifetimes for $M=0.85 M_\odot$
(see data listed in columns 5 and 6 of Table 5).

The rates of MS stars in the three different annuli are --by using the 
star counts based on $B,B-F625W$ (ACS) and $B,B-V$ (WFI) CMDs and the 
mean of the lifetime for Z=0.0002 (2074 Myr) and for Z=0.001 (2211 Myr) --  
the following:  
$19932/2093=9.52\pm0.95\; (r\le r_\alpha)$, 
$19376/2093=9.26\pm0.93\; (r_\alpha < r \le r_\beta)$, 
and $15223/2093=7.27\pm0.73\; (r > r_\beta)$. The errors account for 
uncertainties in both star counts (Poisson) and evolutionary lifetimes 
(10\%). The same outcome applies to the ratios given by star counts 
based on $B,B-F658N$ (ACS) and $B,U-V$ (WFI) CMDs, and indeed we find 
$19935/2093=9.52\pm0.95\; (r\le r_\alpha)$, 
$19815/2093=9.47\pm0.95\; (r_\alpha < r \le r_\beta)$, and 
$14187/2093=6.78\pm0.68\; (r > r_\beta)$.
Therefore, the total rate of MS stars in the selected box is slightly 
more than 25 stars per Myr. This means that on average the HB rate is 
approximately 43\%\footnote{In order to avoid subtle uncertainties 
in the estimate of the relative difference between the HB and the MS rate 
we decided to use $\xi= (r(HB)-r(MS))/[(r(HB)+r(MS))*0.5]$.} larger than 
the MS rate. This discrepancy between 
theory and observations appears as robust empirical evidence, since 
it is marginally affected by the assumed metal abundance. Moreover, 
the ratio between HB and MS stars is also marginally affected by 
field star contamination. Hot HB stars are systematically bluer  
than field stars and the MS stars we selected cover a very narrow 
magnitude range. It is worth mentioning that the RR Lyrae stars 
in \omc ($> 187$, Weldrake, Sackett, \& Bridges 2006, and references therein) 
have not been identified yet in the current photometric catalogue.  
Some of these objects might have a limited coverage of the 
light curves, and therefore they have not been included in 
the sample of EBT1 stars. Therefore, the inclusion of these 
objects implies a mild increase in the observed HB/RG and 
HB/MS star counts ratios.            

This finding and the excess showed by the ratio between the
rate of HB and RG stars when compared to predicted ratios 
(see Fig. 13) is suggesting that we are faced with a true 
excess of HB stars. However, during the last few years several empirical 
and theoretical investigations suggested that a fraction of the order of 
30\% of the stars in \omc might present an increase in He content 
(Bedin et al.\ 2004; Piotto et al.\ 2005; Sollima et al. 2007, and 
references therein). This means 
that before we can reach a firm conclusion on the occurrence of a possible 
discrepancy between predicted and observed star counts we need to investigate 
the impact of the He content on MS, RG, and on HB evolutionary lifetimes.


\section{Dependence on He abundance}

In order to account for the dependence of evolutionary lifetimes on helium 
abundance, we constructed a new series of evolutionary models by assuming 
the same turnoff age ($t\sim 12$ Gyr) and metal abundances, but higher helium 
contents. Fig.~15 shows both hydrogen- and helium-burning evolutionary phases 
for helium-enhanced stellar structures. Note that evolutionary prescriptions 
for helium-burning structures do not cover, as expected, the temperature 
range typical of HB stars. At fixed TO age and metal abundance, an increase 
in the helium content causes a steady decrease in the TO mass. This also means 
that an increase in helium content also causes the HB morphology to become 
systematically bluer (hotter, Caloi, Castellani, \& Tornamb\`e 1978; 
Lee et al. 2005, and references 
therein). At the same time, the He-core mass becomes smaller, in particular for 
Z=0.0002 and Z=0.001 it ranges from 0.514/0.506 (Y=0.23), to 0.494/0.488 (Y=0.33), 
and to 0.479/0.474 (Y=0.42). This means that the extremely hot stellar 
structures are, at fixed effective temperature, fainter (see Figs. 11 and 15), 
and in turn, their lifetimes become systematically longer 
(Castellani et al. 1994; Zoccali et al. 2000). The lifetime of He-enhanced  
EHB structures when compared with canonical (Y=0.23) structures 
increases on average by at least 10\% for Y=0.33 and 20\% for Y=0.42.    

Moreover and even more importantly, the increase in the helium content also 
causes stellar structures around the TO region ($18.65 \le m_B \le 19.15$) to 
have significantly shorter lifetimes (see open and filled circles in Fig. 15).  
In particular, for the metal-poor (Z=0.0002) chemical composition it ranges 
from $t_{MS}\sim 1.42$ Gyr for Y=0.33 to $\sim 0.98$ Gyr for Y=0.42, 
{\it i.e.\/}, 32\% and 53\% shorter when compared with canonical (Y=0.23) 
MS lifetimes. For more metal-rich structures (Z=0.001) the lifetimes range 
from $t_{MS}\sim 1.39$ Gyr for Y=0.33 to $\sim 0.84$ Gyr for Y=0.42, and the 
difference with canonical MS lifetimes is 34\% and 60\%, respectively. 
Note that He-enhanced stellar structures present in this magnitude range 
a very small spread in mass, and indeed $\Delta M \le 0.005 M_\odot$.   
On the other hand, the evolutionary lifetimes along the RGB are only mildly 
shorter (see Figs. 11 and 15), and indeed the difference with canonical RG 
lifetimes is at most of the order of 15\%.

The main outcome of the increase in the He content is that the ratio between 
RG and MS lifetimes increases. Data plotted in the top panel 
of Fig. 16 show that an increase in He abundance of $\sim 45$\% (Y=0.33) 
causes, in the current metallicity range, at least an increase in the 
$t(RG)/t(MS)$ ratio of 25\%, while an increase of $\sim 85$\% (Y=0.42) gives 
an increase of more than a factor of two. The predicted lifetime ratios for 
Y=0.42 and different metal abundances (red and blue dashed lines) attain 
quite similar values over the entire magnitude range.  
This behavior is the result of several nonlinear effects. The variations in
the metal content and in the He abundance independently affect the evolving 
mass at fixed stellar age together with the evolutionary times of MS and RGB
phases. The predicted ratios are systematically higher than observed. However, 
predictions for Z=0.001 and Y=0.33 account for the observed ratio, but 
this would imply that a significant fraction of \omc stars should be metal-rich. 
This evidence is not supported by current metallicity measurements suggesting 
a main peak in the metallicity distribution around $[Fe/H]\sim -1.7$.

Obviously, these ratios only apply if the entire stellar content of \omc 
is He-enriched. In the case where 70\% of the stellar populations presents a 
canonical He content and only 30\% is He-enriched, the predicted ratios 
are significantly smaller (see bottom panel of Fig. 16) and in excellent 
agreement with observed ratios. This outcome applies to the two sets of 
He-enriched models, thus suggesting a marginal dependence on He enhancement. 
A weak dependence of the stellar luminosity function---below the RGB bump---on
helium enhancement has also been recently found by Salaris et al. (2006)
using different sets of cluster isochrones accounting for both the helium
enhancement and the observed abundance anticorrelations in several
heavy elements ($C,N,O,Mg,Al$).  

To further constrain this effect we need to estimate the impact of an enhanced 
He abundance on the ratio of HB stars. Therefore, we use the approach adopted 
in section 4, to estimate the rate of HB stars. The lifetimes typical of the 
three HB groups, by accounting for the two different metal abundances and 
for the same mix of stellar populations are the following: \\

$
t_{EBT1} = 0.70\times t_{EBT1}(Y=0.23) + 0.30\times t_{EBT1}(Y=0.33) = 
     0.70\times [(76+82)\times 0.5] + 0.30\times (81+87)\times 0.5] \sim 81 \;\; Myr 
$

$
t_{EBT2} = 0.70\times t_{EBT2}(Y=0.23) + 0.30\times t_{EBT2}(Y=0.33) = 
     0.70\times [(84+90)\times 0.5] + 0.30\times (90+96)\times 0.5] \sim 89 \;\; Myr 
$  

$
t_{EBT3} = 0.70\times t_{EBT3}(Y=0.23) + 0.30\times t_{EBT3}(Y=0.33) = 
   0.70\times [(94+101)\times 0.5] + 0.30\times (108+117)\times 0.5] \sim 102 \;\;Myr
$  

Only the lifetime of the EBT3 group changes for Y=0.42, since the 
coolest models do not produce HB stars of the EBT1 group and only a 
negligible fraction of HB stars of the EBT2 group.
 
$
t_{EBT3} = 0.70\times t_{EBT3}(Y=0.23) + 0.30\times t_{EBT3}(Y=0.42) = 
  0.70\times [(94+101)\times 0.5] + 0.30\times (123+129)\times 0.5] \sim 106 \;\; Myr
$  

On the basis of these lifetimes, we can now estimate the rates of HB stars. 
In particular for Y=0.33 and for star counts based on $B,B-F625W$ (ACS) 
and $B,B-V$ (WFI) CMDs in the innermost annulus ($ r \le r_\alpha$) we find: \\  

$
r(HB, Y=0.33) = N(EBT3)/102 + N(EBT2)/89 + N(EBT1)/81 
             = 349/102 + 98/89 + 798/81\sim 14.43 
$

while for Y=0.42 we have:\\ 
$
r(HB, Y=0.42) = N(EBT3)/106 + N(EBT2)/87 + N(EBT1)/79 
             = 349/106 + 98/87 + 798/79\sim 14.51  
$

The rate of HB stars for the other annuli and for the star counts based on 
$B,B-F658N$ (ACS) and $B,U-V$ (WFI) CMDs are listed in the last two columns 
of Table 2. Note that an increase in He content causes an increase in the 
HB lifetime, and in turn a mild decrease in the total rate of HB stars.

Fig.~17 shows the ratios of HB and RG rates in the three different annuli. 
These ratios were estimated by accounting for a mix of stellar populations 
with different metal and helium abundances. Interestingly enough, data 
plotted in the top panel show that the predicted ratios for a mix 
of 70\% of \omc stars with canonical helium content and 30\% He-enriched 
(Y=0.42) are systematically larger than observed. 
The discrepancy ranges from at least $\sim 15$\% to $\sim 25$\%, and indeed
the mean in the three different annuli are $1.248\pm 0.046$, $1.177\pm 0.044$,
and $1.183\pm 0.030$. The trend appears similar also for the He-enriched ratios
with Y=0.33, but the discrepancy ranges from $\sim 15$\% to $\sim 20$\%,
and the mean in the three different annuli are $1.207\pm 0.030$, $1.136\pm 0.049$,
and $1.148\pm 0.018$.

To further constrain the mismatch between theory and observations we decided 
to compare the rates of HB and MS stars. The rates of MS star attain, once 
again, very similar values, and indeed in the three different annuli star counts 
based on $B,B-F625W$ (ACS) and $B,B-V$ (WFI) CMDs give: \\ 

$r(MS)=19932/(0.7\times2093+0.30\times 1402.5)=10.57\pm1.10\; (Y=0.33) \;\;\; (r\le r_\alpha)$\\  
\hspace*{1.9truecm} $=19932/(0.7\times2093+0.30\times 910)=11.47\pm1.15\; (Y=0.42)$,  

$r(MS)=19376(0.7\times2093+0.30\times 1402.5)=10.27\pm1.03\; (Y=0.33) \;\;\; (r_\alpha < r \le r_\beta)$\\
\hspace*{1.9truecm} $=19376/(0.7\times2093+0.30\times 910)=11.15\pm1.10\; (Y=0.42)$,    

and 

$r(MS)=15223/(0.7\times2093+0.30\times 1402.5)=8.07\pm0.81\; (Y=0.33) \;\;\; (r > r_\beta)$\\  
\hspace*{1.9truecm} $=15223/(0.7\times2093+0.30\times 910)=8.76\pm0.88\; (Y=0.42) $. 

At the same time, the semi-empirical ratios given by star counts based on 
$B,B-F658N$ (ACS) and $B,U-V$ (WFI) CMDs give:\\

$r(MS)=19935/(0.7\times2093+0.30\times 1402.5)=10.57\pm1.10\; (Y=0.33) \;\;\; (r\le r_\alpha)$\\  
\hspace*{1.9truecm} $=19935/(0.7\times2093+0.30\times 910)=11.47\pm1.15\; (Y=0.42)$, 

$r(MS)=19815/(0.7\times2093+0.30\times 1402.5)=10.51\pm1.05\; (Y=0.33)\;\;\;(r_\alpha < r \le r_\beta)$\\ 
\hspace*{1.9truecm} $=19815/(0.7\times2093+0.30\times 910)=11.40\pm1.14\; (Y=0.42) $,   

and 

$r(MS)=14187/(0.7\times2093+0.30\times 1402.5)=7.52\pm0.75\; (Y=0.33) \;\;\; (r > r_\beta)$\\  
\hspace*{1.9truecm} $=14187/(0.7\times2093+0.30\times 910)=8.16\pm0.81\; (Y=0.42)$. 

Therefore, the total rate of MS stars in the selected box is $\sim 28$ 
stars per Myr for Y=0.33 and $\sim 30$ stars per Myr for Y=0.42. Hence, 
the discrepancy between lifetimes and star counts is of the order of 
30\% for Y=0.33 and of the order of 24\% for Y=0.42. Note that 
the above rates have been estimated by assuming, at fixed helium content,
the mean HB and MS lifetimes for the two different metal abundances 
(Z=0.0002, Z=0.001). 
Current findings indicate that a mix of stellar populations comprising 
70\% with canonical He (Y=0.23) and 30\% with He-enhanced (Y=0.33 
or Y=0.42) stars does not account for the observed excess of HB stars 
in \omcp. 

In passing we note that the HB/MS ratio presents a twofold advantage when 
compared with the RG/MS and the HB/RG ratio: {\em i)} Current predictions 
indicate that an increase in He abundance causes a decrease in the lifetime 
of both RG and MS stars, whereas the lifetime of hot HB stars increases. 
Moreover, the HB/MS ratio has the largest sensitivity to changes in He 
abundance. {\em ii)}  The HB/MS ratio is marginally affected by field 
star contamination (see \S 4).


\section{Discussion and final remarks}

We have presented an extensive photometric investigation on Horizontal Branch, 
Red Giant Branch and Main Sequence stars in the Galactic globular cluster \omcp. 
The central regions of the cluster were covered with a mosaic of
$F435W$, $F625W$, and $F658N-$band data collected with ACS on board 
the Hubble Space Telescope. The outer reaches were covered with a large
set of $U,B,V,I-$band data collected with the mosaic camera, WFI, 
available at the 2.2m ESO/MPI telescope. The current photometric data 
cover more than 99\% of the total flux of the cluster and the final 
catalogue includes approximately 1.7 million stars. This unprecedented
photometric catalogue allowed us to identify more than 3,200 likely
HB stars, the largest sample collected in a globular cluster to date, and
more than 12,500 stars brighter than the base of the RGBs and fainter
than the RGB bumps ($15 \le B \le 18$). The sizable sample of HB stars 
allowed us to constrain on a quantitative basis the change in the HB 
morphology as a function of the radial distance. The relative number of 
extreme HB stars decreases from $\sim 30$\% to $\sim 21$\% when moving 
from the center toward the outer reaches of the cluster. On the other hand, 
the fraction of less hot HB stars (EBT1) increases from $\sim 62$\% to $\sim 72$\%. 

Two physical mechanisms that might partially account for the discrepancy between
observed and predicted star counts are the so-called "breathing pulses" (Castellani
et al. 1985) and mild convective core overshooting. These mechanisms increase the 
central He-burning lifetime by roughly the 15\% (Caputo et al. 1989; 
Cassisi et al. 2001; Straniero et al. 2003).
On the other hand, stellar rotation has an opposite trend because the rotation delays
the He core flash at the tip of the RGB. This causes an increase in the mass of the
He core and a decreases in the envelope mass (Renzini 1977), since the mass loss is
more efficient close to the tip of the RGB. Therefore, rotating HB structures are
typically bluer and brighter than ``canonical" HB stars. Moreover, one should also
take into account the rotationally induced mixing along the RGB phase (Denissenkov
\& VandenBerg 2003; Palacios et al. 2006). If the envelope mixing does not reach
the edge of the H-burning shell the effect on HB models is negligible. On the contrary,
if the mixing reaches the H-burning shell, then the He abundance in the envelope of
bright RGB structures increases, thus causing a higher luminosity and a larger mass
loss close to the tip of RGB. The aftermath of such a mechanism is a brighter and
bluer HB morphology (see e.g. Sweigart \& Catelan 1998, and references therein).
The rotational HB models present, when compared with canonical ones, larger He core
masses, and in turn more luminous stellar structures. This means that the HB lifetime
of the rotational HB models is, at fixed effective temperature, shorter than for
canonical ones. Moreover, empirical evidence suggests that the projected rotational
velocity among hot HB stars ($T_e > 10,000$ K), in several GCs that show a well-defined
Blue Tail, is quite constant and smaller than 10 km/sec (see Fig. 5 in Recio Blanco
et al. 2004), while it presents a mild increase only among cooler ($T_e \le 9,000$ K)
HB stars (Behr et al. 2000).

This evidence might suggest that the higher density in the central
regions plays a crucial role in the formation of extreme HB stars.
However, \omc presents a central density that is at least one order of
magnitude smaller ($\log \rho_0 =3.12 \; L_\odot pc^{-3}$) than other
massive GGCs (47 Tuc, NGC~6397). Therefore, it might be interesting to
estimate the expected frequency of stellar collisions for this very massive
GC. It was demonstrated by King, Surdin \& Rastorguev (2002) that the number
of stellar collisions per year for a GC with a King profile is given by
$\Gamma_c = 5\times10^{-15}\, (\Sigma_0^3\times r_c)^{1/2}$, where $r_c$ is
the core radius in parsec and $\Sigma_0$ is the central surface brightness
in units of $L_{\odot,V} pc^{-2}$. To estimate the $r_c$ in parsec we
adopted the true distance modulus ($\mu=13.70\pm0.6\pm0.6$) recently provided
by Del Principe et al. (2006) and the core radius ($r_c\sim 2.6$ arcmin)
provided by Trager et al. (1995) and more recently by Ferraro et al. (2006).
We find $r_c=4.13$ pc, and in turn that $\Gamma_c$, i.e. the number of
collisions is $5.85\times 10^{-9} yr^{-1}$. Following the assumptions adopted
by Piotto et al. (2004) in order to estimate the probability that a single star
undergoes a stellar collision in 1 yr, $\Gamma_\star$, we divided  $\Gamma_c$ by
the number of cluster stars. In order to estimate $N_\star$ we adopted the
mass-to-light ratio ($M/L_V\sim2.5$) recently estimated by van de Ven et al. (2006),
the absolute visual magnitude of \omc is $M_V=-10.39$ ($V_t=3.68$, $E(B-V)=0.12$,
Harris 1996) and a typical mass of $0.4 M_\odot$. The total number of stars we find
is $7.74\times10^6$. This estimate agrees quite well with the dynamical
estimate of the total mass of the cluster provided by van de Ven et al. (2006).
They found $M_t=2.5\times 10^6\; M_\odot$ that assuming the same typical
mass provides a number of stars in \omc of $6.25\times10^6$. Therefore, the
probability that a single star centrally located experiences a collision in 1 yr is
$\log \Gamma_\star=-15.12$. The current collision rate is approximately one order
of magnitude smaller than the collision rate in NGC2808 ($\log \Gamma_\star=-14.05$)
a relatively massive GC ($M_V=-9.36$, Harris 1996) characterized by a higher central
density ($\log \rho_0 =4.61 \; L_\odot pc^{-3}$) and a HB morphology including a
significant number of red HB stars and very extended blue HB tail
(Castellani et al. 2006a). However, the current collision rate in \omc is slightly
higher than the collision rate of NGC~4833 ($\log \Gamma_\star=-15.31$) a GC with
a central density ($\log \rho_0 =3.06 \; L_\odot pc^{-3}$) and a HB morphology
(Melbourne et al. 2000) quite similar to \omcp.
Current findings seem to support the evidence brought forward by
Castellani et al. (2006a) that the Blue Tails, if affected by cluster dynamics,
should be considered more a transient phenomenon rather than an intrinsic feature
of GCs. We plan to address this topic on the basis of a larger sample of GCs in a
forthcoming paper.

The referee suggested two different working hypotheses that might account for the
peculiar radial distribution of HB stars.\\
{\em i}) Let us assume that extreme HB stars are the progeny of WD binaries.
This would imply that the radial gradient of HB stars could be the aftermath of
mass segregation causing a larger fraction of binaries to be located in the
cluster center. Recent empirical evidence indicates that more massive GCs tend
to have a larger number of HB stars and bluer HB morphologies (Davies et al. 2004;
Recio Blanco et al. 2006). However, Ferraro et al. (2006) found a lack of mass
segregation in \omc, since the radial distribution of Blue Stragglers does not
appear to be centrally peaked. These objects are thought to be the evolution of
binary systems and their radial distribution in GCs is typically more concentrated
than single cluster stars.\\
{\em ii}) Let us assume that a radial gradient in the initial He abundance
was already present at the epoch of the cluster formation. Could this initial
gradient still persist due to the long two-body relaxation time of \omc?
This working hypothesis is supported by the empirical evidence that more
metal-rich cluster stars are more centrally concentrated than metal-poor
ones (Norris et al. 1996, 1997). Moreover, Sollima et al. (2007) found
that stars belonging to the so-called "blue-MS" are more centrally
concentrated than stars belonging to the "red-MS". These findings together
with the evidence that stars in the "blue-MS" appear to be less metal-poor
(Piotto et al. 2005) and that the two body relaxation time increases outside
the half-mass radius suggest that a radial gradient in He abundance may have 
been set up at the epoch of cluster formation. However, recent chemical 
evolution models might account for the metallicity distribution in \omc, 
but do not account for the significant He enhancement suggested to explain 
the "blue-MS" (Romano et al. 2007).

To provide firm constraints on the stellar populations in \omc we performed 
detailed comparisons of large samples of evolved and MS stellar tracers 
with homogeneous evolutionary prescriptions. In particular, the comparison 
between the observed star counts (RG, HB, MS) and the theoretical framework 
based on a canonical helium content (Y=0.23) brings forward several interesting 
findings.  

{\em i) Star counts along the RGB -} Evolutionary lifetimes for stellar structures 
along the RGB phase constructed by adopting a broad range of metal abundances 
(Z=0.0002, 0.001), different stellar masses ($M=0.80, 0.85 M_\odot$) agree
quite well with empirical RG star counts fainter the RGB Bump.   

{\em ii) Stellar population ratios-} The comparison between theory and observations 
discloses that the observed ratio of HB and RG stars is systematically larger
(30\% -- 40\%) than the ratio of HB and RG lifetimes predicted by evolutionary 
models with metal abundances (Z=0.0002, Z=0.001) that bracket the observed spread 
in metallicity of \omc stars.  

{\em iii) The ratio between HB, RG, and MS stars-} The comparison between the observed 
ratio of RG and MS stars with the predicted ratio of their evolutionary lifetimes is 
suggestive of a mild excess of RG stars in the faint region of the RGB. By using  
evolutionary tracks with canonical compositions, the discrepancy ranges once again 
from 10 to 15\%. The same outcome applies to the observed ratio between HB and MS 
stars. However, their excess when compared with the predicted ratio is of the 
order of 43\%. This circumstantial evidence suggests the possibility that a 
fraction of bright RG stars miss the central helium flash evolutionary phase.\\ 

Recent empirical evidence concerning the 
occurrence in \omc of a blue MS and of a well-defined blue HB tail have been 
explained with the presence of a He-enhanced stellar population. To constrain 
this appealing working hypothesis we constructed several evolutionary tracks and 
HB evolutionary models using the same metallicity range but higher He contents, 
namely Y=0.33 and Y =0.42. The comparison between theory and observations was 
performed by assuming a mix of stellar populations made, according to empirical 
evidence, with 70\% of canonical stars and 30\% He-enhanced stars. We found 
the following results.  

{\em i) The ratio between RG, HB, and MS stars-} The observed ratio of RG and 
MS stars agrees quite-well with the predicted lifetimes of mixed He-enhanced 
(Y=0.33 or Y =0.42) stellar populations. Thus suggesting a negligible 
sensitivity of this ratio on the He content. 
On the other hand, the observed HB and RG star counts are still 15\%--25\% larger
than predicted for He-mixed stellar populations with (Y=0.42), but the discrepancy
is on average a factor of two smaller than for predicted ratios based on evolutionary
models with canonical He abundance. The same outcome applies to the predicted ratio
for Y=0.33 but the discrepancy ranges from $\sim$ 15\% to $\sim$ 20\%. 
Moreover and even more importantly, the observed ratios between HB and MS 
star counts are once again from $\sim$ 24\% and $\sim$ 30\% larger 
than predicted for He-mixed populations with Y=0.42 and Y=0.33, respectively.     

{\em ii) New empirical constraints-} The findings based on standard and He-mixed
stellar populations have quite different impacts on the final evolutionary fate 
of low-mass stars. The theoretical scenario based on a canonical helium content 
also implies that a fraction of RG stars does not approach the tip of the 
RGB, missing the core-helium flash phase, and might end up their evolution 
{\em either} as extreme HB stars {\em or} as He-core WDs (Castellani et al. 2006b). 
Therefore, we should also observe an excess of AGB{\em-manqu\`e} stars and of 
bright WDs. The theoretical framework based on He-mixed stellar populations 
is puzzling, and indeed the ratios between HB and MS stars is roughly a factor 
of two larger than the excess between HB and RG stars. However, the ratios between 
MS and RG star counts agree quite-well with the ratio of evolutionary lifetimes.    

During the last few years several homogeneous photometric investigations covering
a substantial fraction of the body of GGCs found mounting evidence for peculiar 
radial distributions among different stellar tracers (Castellani et al. 2006a; 
Sollima et al. 2007). Moreover, a significant excess of RG stars above 
and below the RGB bump has also been detected in several GCs (Sandquist 
et al. 1999; Pollard et al. 2005). On the other hand, in a very recent 
investigation Sandquist \& Martel (2006) found a well-defined deficiency 
of RG stars along the RGB of the massive GGC NGC2808. This globular shows 
an extended blue HB tail and the authors also suggest that strong mass 
loss events might account for such an empirical evidence. 
An accurate and quantitative analysis of RG stars in \omc across and above 
the RGB bump is mandatory to constrain the efficiency of this possible 
evolutionary chanel. It goes without saying that the new mosaic CCD cameras 
with FOV of the order of 1 square degree might play a crucial role to provide 
sound quantitative constraints on these new compelling features.      


\acknowledgements
It is a pleasure to thank S. Cassisi for several discussions and a 
detailed reading of an early draft of this manuscript. We also thank 
F. De Angeli for many enlightening suggestions and insights concerning 
the collisional rate of GCs. We acknowledge an anonymous referee for 
his/her positive and pertinent comments that helped us to improve the 
readability of the manuscript. 
This work was partially supported by PRIN-INAF2005 (P.I.: A. Buzzoni), 
"Galactic Stellar Populations", by PRIN-INAF2004 (P.I.: M. Bellazzini), 
"A hierarchical merging tale told by stars: motions, ages and chemical
compositions within structures and substructures of the Milky Way", 
and by Particle Physics and Astronomy Research Council (PPARC). 
We also thank the ESO and the HST Science Archive for their prompt support.
This publication makes use of data products from VizieR (Ochsenbein, Bauer, 
\& Marcout 2000) and from the Two 
Micron All Sky Survey, which is a joint project of the University of Massachusetts 
and the Infrared Processing and Analysis Center/California Institute of
Technology, funded by the National Aeronautics and Space Administration
and the National Science Foundation.


\clearpage
\tablewidth{0pt}
\begin{deluxetable}{lccrrcc}
\tabletypesize{\scriptsize}
\tablecaption{Log of scientific CCD images of Omega Centauri collected with the 
WFI available at the 2.2m ESO/MPI telescope.\tablenotemark{a}}\label{tbl-1}
\tablehead{
\colhead{Frame\tablenotemark{b}}&
\colhead{RA\tablenotemark{c}}&
\colhead{DEC\tablenotemark{d}}&
\colhead{MJD\tablenotemark{e}}&
\colhead{Exposure\tablenotemark{f}}&
\colhead{Filter\tablenotemark{g}}&
\colhead{Seeing\tablenotemark{h}} \\
\colhead{(1)}&
\colhead{(2)}&
\colhead{(3)}&
\colhead{(4)}&
\colhead{(5)}&
\colhead{(6)}&
\colhead{(7)} } 
\startdata
WFI.1999-01-24T08:18:52.934 & 13:26:18.9 & -47:36:01.1 & 51202.3464460 & 29.92  & B  & 1.11\\ 
WFI.1999-01-24T08:21:57.925 & 13:26:42.7 & -47:36:01.5 & 51202.3485871 & 29.92  & B  & 0.91\\ 
WFI.1999-01-24T08:24:01.313 & 13:26:48.6 & -47:36:01.8 & 51202.3500152 & 29.92  & B  & 1.00\\
WFI.1999-01-24T08:25:44.311 & 13:26:48.7 & -47:36:01.5 & 51202.3512073 & 299.92 & B  & 0.98\\
WFI.1999-03-25T05:28:39.461 & 13:26:44.8 & -47:26:34.2 & 51262.2282345 & 149.92 & I  & 0.84\\
WFI.1999-03-25T05:32:42.409 & 13:26:47.8 & -47:26:03.8 & 51262.2310464 & 149.92 & I  & 0.75\\
WFI.1999-03-25T05:36:40.398 & 13:26:50.8 & -47:25:34.4 & 51262.2338009 & 149.92 & I  & 0.73\\
WFI.1999-03-25T06:04:59.817 & 13:26:44.8 & -47:26:34.5 & 51262.2534701 & 199.92 & V  & 0.86\\
WFI.1999-03-25T06:10:25.156 & 13:26:47.8 & -47:26:04.6 & 51262.2572356 & 199.92 & V  & 0.95\\
WFI.1999-03-25T06:14:59.839 & 13:26:50.7 & -47:25:33.7 & 51262.2604148 & 199.92 & V  & 1.02\\
WFI.1999-04-29T00:19:41.952 & 13:26:46.4 & -47:35:48.5 & 51297.0136800 & 89.92  & I  & 0.59\\
WFI.1999-04-29T00:23:20.786 & 13:26:46.4 & -47:31:38.8 & 51297.0162128 & 89.92  & I  & 0.59\\
\enddata                                                                                     
\tablenotetext{a}{Table 1 is available in its entirety via the link to the 
machine-readable version above. A portion is shown here for guidance regarding 
its form and content.}                                                               

\tablenotetext{b}{Frame identification.}                                                                     
\tablenotetext{c}{Right ascension (J2000) in units of hours, minutes, and seconds.}                                                                  
\tablenotetext{d}{Declination (J2000) in units of degrees, arcminutes, and 
arcseconds.}                                                                      
\tablenotetext{e}{Modified Julian Date (JD - 2400000.5) at the start of the 
exposure.}                                                  
\tablenotetext{f}{Exposure time (sec).} 
\tablenotetext{g}{Filter name. $U-$band and $B-$band data collected in 1999 and  
2000 observing runs adopted the filters $U/38 (ESO\#841)$ and $B/99 (ESO\#842)$, 
while subsequent data have been collected with the new filters 
$U/50 (ESO\#877)$ and $B/123 (ESO\#878)$.} 
\tablenotetext{h}{Seeing of individual images (arcsec).} 
\end{deluxetable}
                                                                             
\clearpage

\begin{deluxetable}{l r r r r r r r}
\tabletypesize{\scriptsize}
\tablecaption{
The number of HB stars in the three selected HB subgroups detected 
in the three different regions of the cluster. Figures within
brackets give for each region the relative fraction of the 
different subgroups with respect to the total number of HB stars.
\label{tab_hb_01}}
\tablewidth{0pt}
\tablehead{
\colhead{Radius}  &
\colhead{N(EBT1)}  &
\colhead{N(EBT2)}  &
\colhead{N(EBT3)}  &
\colhead{$N_{tot}$}& 
\colhead{$r(HB)$\tablenotemark{a}} & 
\colhead{$r(HB)$\tablenotemark{b}} & 
\colhead{$r(HB)$\tablenotemark{c}}
}
\startdata
\multicolumn{8}{c}{Star counts based on $B,B-F625W$ (ACS) and $B,B-V$ (WFI) CMDs}\\

$ r \le r_\alpha$           & 798 (\it $64\pm3$\%)    & 98 (\it $8\pm1$\%)  & 349 (\it $28\pm2$\%)    & 1245 & 14.79 & 14.43 & 14.51 \\

$ r_\alpha < r \le r_\beta$ & 402\tablenotemark{d}+293\tablenotemark{e} (\it $60\pm3$\%)& 56\tablenotemark{d}+40\tablenotemark{e} (\it $8\pm1$\%) & 235\tablenotemark{d}+136\tablenotemark{e} (\it $32\pm2$\%)& 1162 & 13.69 & 13.34  & 13.39 \\

$ r> r_\beta$               & 650 (\it $72\pm4$\%)    & 64 (\it $7\pm1$\%)   & 190 (\it $21\pm2$\%)    & 904 & 10.90 & 10.65& 10.75\\
Total                       & 2143 (\it $65\pm2$\%)   & 278 (\it $8\pm1$\%)  & 910 (\it $27\pm1$\%)    & 3311& 39.38 & 38.42& 38.64\\

\multicolumn{8}{c}{Star counts based on $B,B-F658N$ (ACS) and $B,U-V$ (WFI) CMDs}\\

$ r \le r_\alpha$           & 787 (\it $64\pm3$\%)    & 99 (\it $8\pm1$\%)  & 349 (\it $28\pm2$\%)    & 1235 & 14.66 & 14.30  & 14.38 \\

$ r_\alpha < r \le r_\beta$ & 404\tablenotemark{d}+299\tablenotemark{e} (\it $61\pm3$\%)& 59\tablenotemark{d}+40\tablenotemark{e} (\it $9\pm1$\%) & 237\tablenotemark{d}+110\tablenotemark{e} (\it $30\pm2$\%)& 1149 & 13.58  & 13.24 &  13.30\\

$ r> r_\beta$               & 618 (\it $72\pm4$\%)    & 61 (\it $7\pm1$\%)   & 181 (\it $21\pm2$\%)    & 860 & 10.37 & 10.13 & 10.23 \\
Total                       & 2108 (\it $65\pm2$\%)   & 269 (\it $8\pm1$\%)  & 877 (\it $27\pm1$\%)    & 3244 & 38.61& 37.67 & 37.90 \\
\enddata
\tablenotetext{a}{Rate of HB stars, {\it i.e.\/}, the number of HB stars formed per Myr according to the lifetime of 
the three different EBT groups. The lifetimes were estimated as the mean of HB models with Z=0.0002 (Y=0.23) 
and Z=0.001 (Y=0.232).}
\tablenotetext{b}{Rate of HB stars, {\it i.e.\/}, the number of HB stars formed per Myr according to the lifetime of
the three different EBT groups. The lifetimes were estimated by accounting for a mix of stellar populations with different 
metal (Z=0.0002, 0.001) and helium (Y=0.23, 0.33) abundances.}
\tablenotetext{c}{Rate of HB stars, {\it i.e.\/}, the number of HB stars formed per Myr according to the lifetime of
the three different EBT groups. The lifetimes were estimated by accounting for a mix of stellar populations with different 
metal (Z=0.0002, 0.001) and helium (Y=0.23, 0.42) abundances.}
\tablenotetext{d}{Star counts based on ACS data ($r_\alpha < r \le r_{acs}$).}
\tablenotetext{e}{Star counts based on WFI data ($r_{acs} < r \le r_\beta$).}
\end{deluxetable}

\clearpage


\begin{deluxetable}{l r r r r }
\tabletypesize{\scriptsize}
\tablecaption{
The number of RG stars detected in the three selected cluster regions.
Figures within brackets give for each region the relative fraction   
with respect to the total number of RG stars.
\label{tab_hb_02}}
\tablewidth{0pt}
\tablehead{
\colhead{Radius}  &
\colhead{$RG$\tablenotemark{a}}  &
\colhead{$\omega3$\tablenotemark{a}}  &
\colhead{$RG$\tablenotemark{b}}  &
\colhead{$\omega3$\tablenotemark{b}}  
}
\startdata
$ r \le r_\alpha$           &  4527      & 138 (\it $3.0\pm0.26$\%)  & 4598      &  153  (\it $3.0\pm0.25$\%) \\

$ r_\alpha < r \le r_\beta$ &  2469\tablenotemark{c}+1840\tablenotemark{d} & 67\tablenotemark{c}+103\tablenotemark{d}  (\it $4.0\pm0.31$\%)   & 2504\tablenotemark{c}+1913\tablenotemark{d} &  77\tablenotemark{c}+67\tablenotemark{d} (\it $3.0\pm0.25$\%) \\ 

$ r> r_\beta$               &  3492      &  286 (\it $8.0\pm0.5$\%)    & 3209      & 195  (\it $6.0\pm0.44$\%)\\ 
Total                       & 12328      &  594 (\it $5.0\pm0.21$\%)    & 12224     & 492  (\it $4.0\pm0.18$\%)\\ 
\enddata
\tablenotetext{a}{Star counts based on $B,B-F625W$, $B,B-V$ CMDs.}
\tablenotetext{b}{Star counts based on $B,B-F658N$, $B,U-V$ CMDs.}
\tablenotetext{c}{Star counts based on ACS data ($r_\alpha < r \le r_{acs}$).}
\tablenotetext{d}{Star counts based on WFI data ($r_{acs} < r \le r_\beta$).}
\end{deluxetable}

\clearpage


\begin{deluxetable}{c c c c c c c}
\tabletypesize{\scriptsize}
\tablecaption{Empirical star counts in different "windows" along the RGB. 
Star counts have been performed in different CMDs. 
\label{tab_hb_03}}
\tablewidth{0pt}
\tablehead{
\colhead{$\Delta m_B$}  &
\colhead{$r \le r_\alpha$\tablenotemark{a}}  &
\colhead{$r_\alpha < r \le r_\beta$\tablenotemark{a}}  &
\colhead{$ r> r_\beta$\tablenotemark{a}}  &  
\colhead{$r \le r_\alpha$\tablenotemark{b}}  &
\colhead{$r_\alpha < r \le r_\beta$\tablenotemark{b}}  &
\colhead{$ r> r_\beta$\tablenotemark{b}}    
}
\startdata
17.6--16.6 &  2413  & 2221\tablenotemark{c}  & 1837 & 2450 & 2277\tablenotemark{c} & 1668 \\  
17.5--16.5 &  2187  & 2060  & 1705 & 2228 & 2111 & 1555 \\
17.4--16.4 &  2016  & 1890  & 1544 & 2061 & 1953 & 1408 \\
17.3--16.3 &  1842  & 1750  & 1411 & 1885 & 1793 & 1294 \\ 
17.2--16.2 &  1639  & 1584  & 1285 & 1685 & 1627 & 1177 \\ 
17.1--16.1 &  1480  & 1448  & 1184 & 1515 & 1489 & 1081 \\  
17.0--16.0 &  1369  & 1340  & 1072 & 1405 & 1375 & ~~988 \\
16.9--15.9 &  1241  & 1246  & ~~983  & 1269 & 1278 & ~~903 \\
16.8--15.8 &  1138  & 1168  & ~~903  & 1164 & 1202 & ~~835 \\
16.7--15.7 &  1084  & 1133  & ~~828  & 1103 & 1169 & ~~774 \\ 
16.6--5.6 &  1010  & 1058  & ~~787  & 1030 & 1083 & ~~738 \\ 
\enddata
\tablenotetext{a}{ Star counts based on $B,B-F625W$, $B,B-V$ CMDs.
}
\tablenotetext{b}{ Star counts based on $B,B-F658N$, $B,U-V$ CMDs.
}
\tablenotetext{c}{ Star counts account for both ACS and WFI data.
}
\end{deluxetable}

\clearpage


\begin{deluxetable}{c c c c c c}
\tabletypesize{\scriptsize}
\tablecaption{
Theoretical predictions concerning the different lifetimes that a stellar 
structure with canonical He abundance ($Y\sim 0.23$), but different stellar 
masses and chemical compositions spent in the different "windows" along the RGB. 
The numbers in parentheses are power of ten.  
\label{tab_hb_04}}
\tablewidth{0pt}
\tablehead{
\colhead{$\Delta m_B$}  &
\colhead{$\Delta M_B$\tablenotemark{a}}  &
\colhead{$\Delta t$\tablenotemark{b}}  &
\colhead{$\Delta t$\tablenotemark{b}}  &  
\colhead{$\Delta t$\tablenotemark{c}}  &
\colhead{$\Delta t$\tablenotemark{c}}  \\  
\colhead{(mag)}  &
\colhead{(mag)}  &
\colhead{(Z=0.0002)}  &
\colhead{(Z=0.001)}   & 
\colhead{(Z=0.0002)}  &
\colhead{(Z=0.001)}     
}
\startdata
 17.60--16.61 &  3.44-2.44 &  2.33(8) &  2.08(8) &  2.45(08) & 2.01(8) \\  
 17.50--16.51 &  3.34-2.35 &  2.13(8) &  1.90(8) &  2.05(08) & 1.86(8) \\ 
 17.40--16.40 &  3.24-2.24 &  1.97(8) &  1.76(8) &  1.82(08) & 1.74(8) \\ 
 17.30--16.30 &  3.14-2.14 &  1.81(8) &  1.62(8) &  1.70(08) & 1.60(8) \\  
 17.20--16.20 &  3.04-2.04 &  1.68(8) &  1.49(8) &  1.58(08) & 1.47(8) \\ 
 17.10--16.11 &  2.94-1.95 &  1.55(8) &  1.37(8) &  1.49(08) & 1.36(8) \\  
 17.00--16.00 &  2.84-1.84 &  1.43(8) &  1.27(8) &  1.39(08) & 1.23(8) \\  
 16.90--15.91 &  2.74-1.75 &  1.31(8) &  1.16(8) &  1.29(08) & 1.14(8) \\  
 16.80--15.80 &  2.64-1.64 &  1.21(8) &  1.05(8) &  1.18(08) & 1.06(8) \\  
 16.70--15.70 &  2.54-1.54 &  1.11(8) &  9.75(7) &  1.08(08) & 9.76(7) \\ 
 16.60--15.62 &  2.44-1.45 &  1.01(8) &  8.96(7) &  1.00(08) & 8.91(7) \\  
\enddata
\tablenotetext{a}{Absolute $B-$band magnitudes for stellar structures with 
$M=0.80 M_\odot$, Z=0.0002, and Y=0.23.}
\tablenotetext{b}{Stellar ages (years) for stellar structures with 
$M=0.80 M_\odot$ and different chemical compositions.}
\tablenotetext{c}{Stellar ages (years) for a stellar structure with 
$M=0.85 M_\odot$ and different chemical compositions.}
\end{deluxetable}

\clearpage


\begin{deluxetable}{c c c c c c}
\tabletypesize{\scriptsize}
\tablecaption{
Theoretical predictions concerning the different lifetimes that a stellar 
structure at fixed mass value but with different metal abundances and a 
He-enhanced (Y=0.33, 0.42) composition spend in the different "windows" 
along the RGB. The numbers in parentheses are power of ten.
\label{tab_hb_05}}
\tablewidth{0pt}
\tablehead{
\colhead{$\Delta m_B$}  &
\colhead{$\Delta M_B$\tablenotemark{a}}  &
\colhead{$\Delta t$\tablenotemark{b}}  &
\colhead{$\Delta t$\tablenotemark{b}}  &  
\colhead{$\Delta t$\tablenotemark{c}}  &
\colhead{$\Delta t$\tablenotemark{c}}  \\  
\colhead{(mag)}  &
\colhead{(mag)}  &
\colhead{(Z=0.0002)}  &
\colhead{(Z=0.001)}   & 
\colhead{(Z=0.0002)}  &
\colhead{(Z=0.001)}     
}
\startdata
17.61--16.60 & 2.90-1.84 & 2.17(8) &  1.78(8) & 2.16(8) &  1.77(8) \\
17.50--16.50 & 2.78-1.73 & 1.97(8) &  1.66(8) & 1.91(8) &  1.62(8) \\
17.41--16.41 & 2.69-1.63 & 1.84(8) &  1.53(8) & 1.78(8) &  1.50(8) \\
17.31--16.31 & 2.58-1.52 & 1.70(8) &  1.41(8) & 1.64(8) &  1.39(8) \\
17.21--16.21 & 2.48-1.42 & 1.58(8) &  1.31(8) & 1.48(8) &  1.28(8) \\
17.11--16.11 & 2.37-1.31 & 1.45(8) &  1.20(8) & 1.35(8) &  1.18(8) \\
17.01--16.00 & 2.27-1.20 & 1.35(8) &  1.10(8) & 1.25(8) &  1.08(8) \\
16.91--15.90 & 2.16-1.10 & 1.23(8) &  1.01(8) & 1.15(8) &  9.78(7) \\
16.81--15.80 & 2.06-0.99 & 1.14(8) &  9.24(7) & 1.05(8) &  8.91(7) \\
16.70--15.70 & 1.94-0.88 & 1.03(8) &  8.45(7) & 9.64(7) &  8.22(7) \\
16.60--15.60 & 1.84-0.77 & 9.47(7) &  7.75(7) & 8.72(7) &  7.60(7) \\
\enddata
\tablenotetext{a}{Absolute $B-$band magnitudes for stellar structures with 
$M/M_\odot=0.655$, Z=0.0002, and Y=0.33.}
\tablenotetext{b}{Stellar ages (years) for helium enhanced (Y=0.33) 
stellar structures constructed at fixed TO-age (12 Gyr) and metal 
abundance. The TO-mass are $M/M_\odot=0.655$, for Z=0.0002 and 
$M/M_\odot=0.665$ for Z=0.001.}
\tablenotetext{c}{Stellar ages (years) for helium enhanced (Y=0.42) 
stellar structures constructed at fixed TO-age (12 Gyr) and metal 
abundance. The TO-mass is $M/M_\odot=0.55$, for both Z=0.0002 and 
and Z=0.001 structures.}
\end{deluxetable}


\clearpage
\begin{figure}[!h]
\begin{center}
\includegraphics[height=0.65\textheight,width=0.80\textwidth]{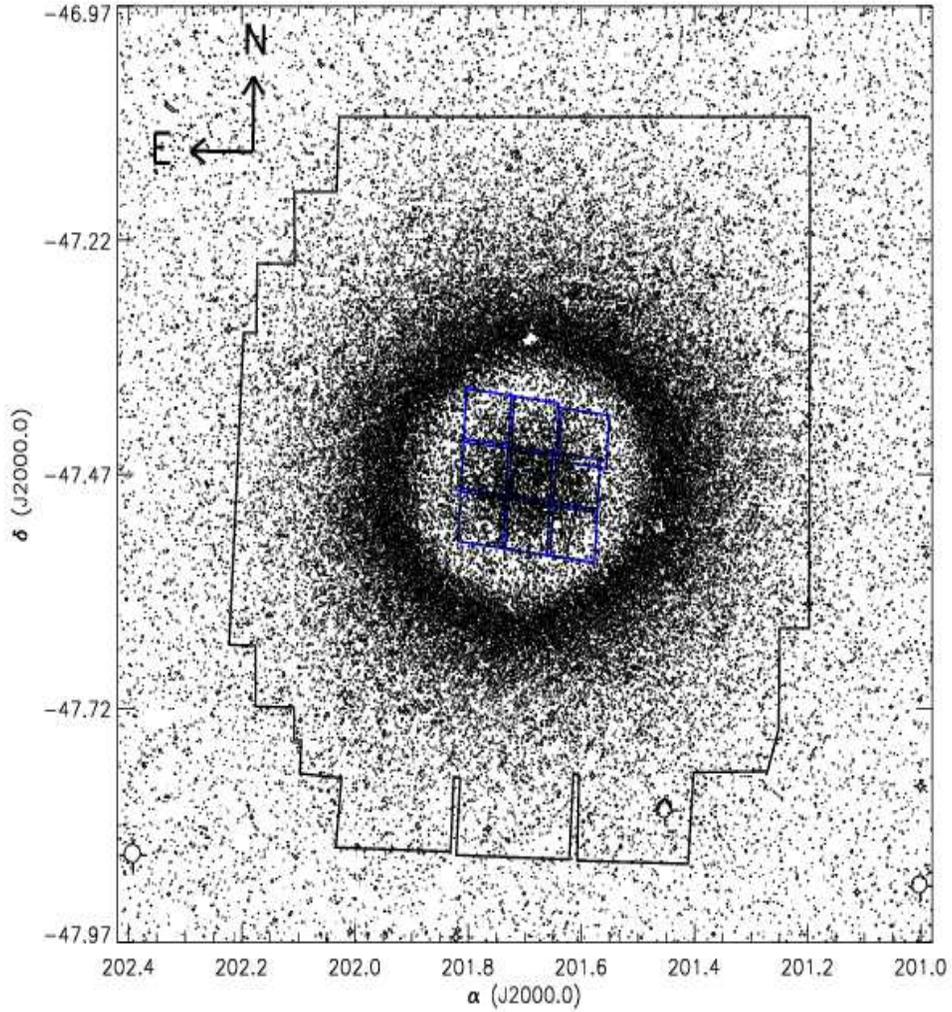}
\caption{Location of the pointings of the ground-based $U,B,V,I-$band data 
collected with WFI@2.2m ESO/MPI (black line) and the space 
$F435W, F625W, F658N-$band data collected with ACS@HST (blue mosaic). 
The outline of individual WFI pointings is plotted over a $1\times1$ square 
degree field in the Digital Sky Survey (DSS). The central cluster field, 
unresolved by the DSS, is superposed by a sample from our star catalogue 
to show the location of the HST mosaic.   
}\label{fig1}
\end{center}
\end{figure}

\clearpage
\begin{figure}[!h]
\begin{center}
\includegraphics[height=0.65\textheight,width=0.80\textwidth]{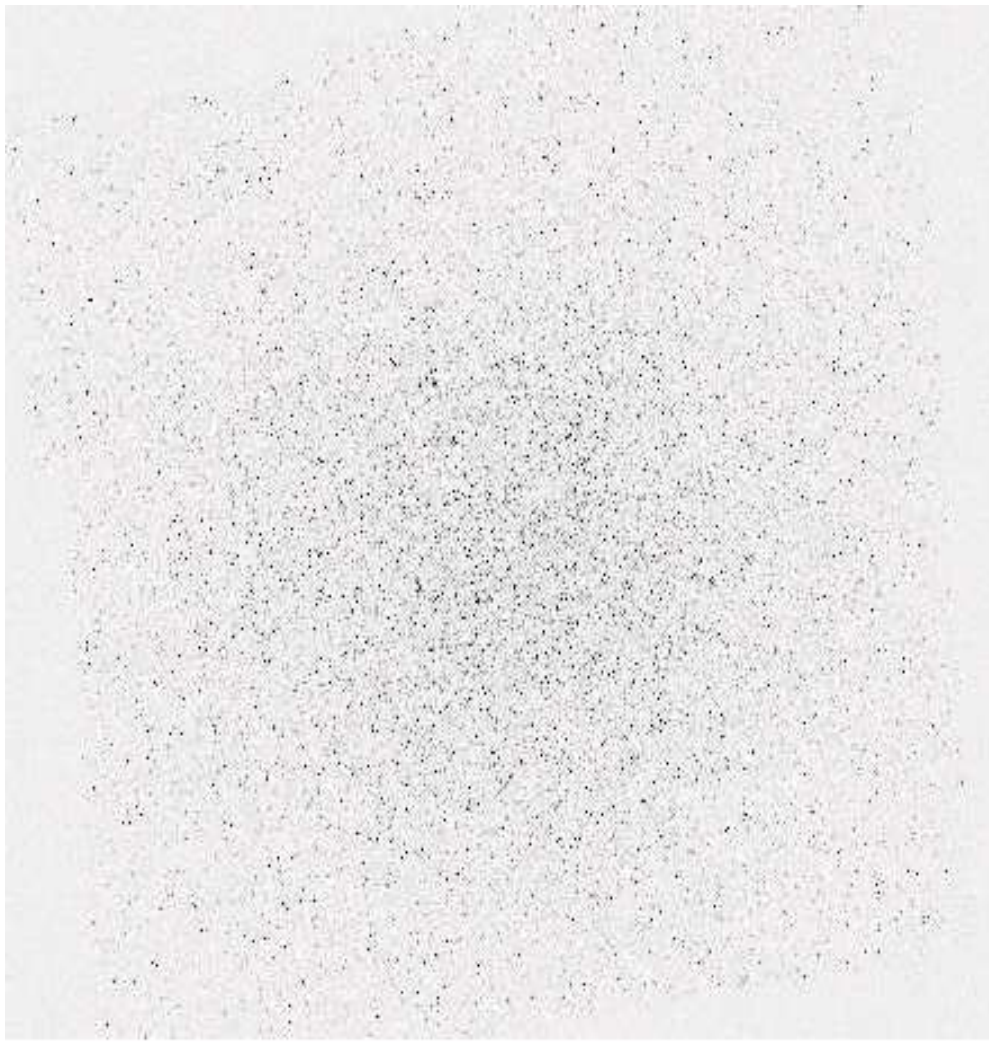}
\caption{
The color image is a true color hybrid composition of WFI and ACS images. 
The catalogue based on ACS images provides the spatial information, while 
the colors of individual stars are based on $U,B,V,I-$band WFI data. 
In order to preserve the appropriate resolution three artificial images 
of $8192\times8192$ pixels have been created: one for the combined 
$U$ and $B-$band data, and two for the $V$ and the $I-$band data. 
Approximately 1 million stars have been included 
in each of these images using the {\tt artdata.mkobjects} task, resulting 
in an image with almost the same pixel scale as an ACS image. The three 
independent channels have then been merged to create a true-color image.
The mosaic's field on the sky is $\approx 10\arcmin \times 9.5\arcmin$.
north is up, and east is to the left. 
}\label{fig2}
\end{center}
\end{figure}

\clearpage
\begin{figure}[!h]
\begin{center}
\includegraphics[height=0.65\textheight,width=0.80\textwidth]{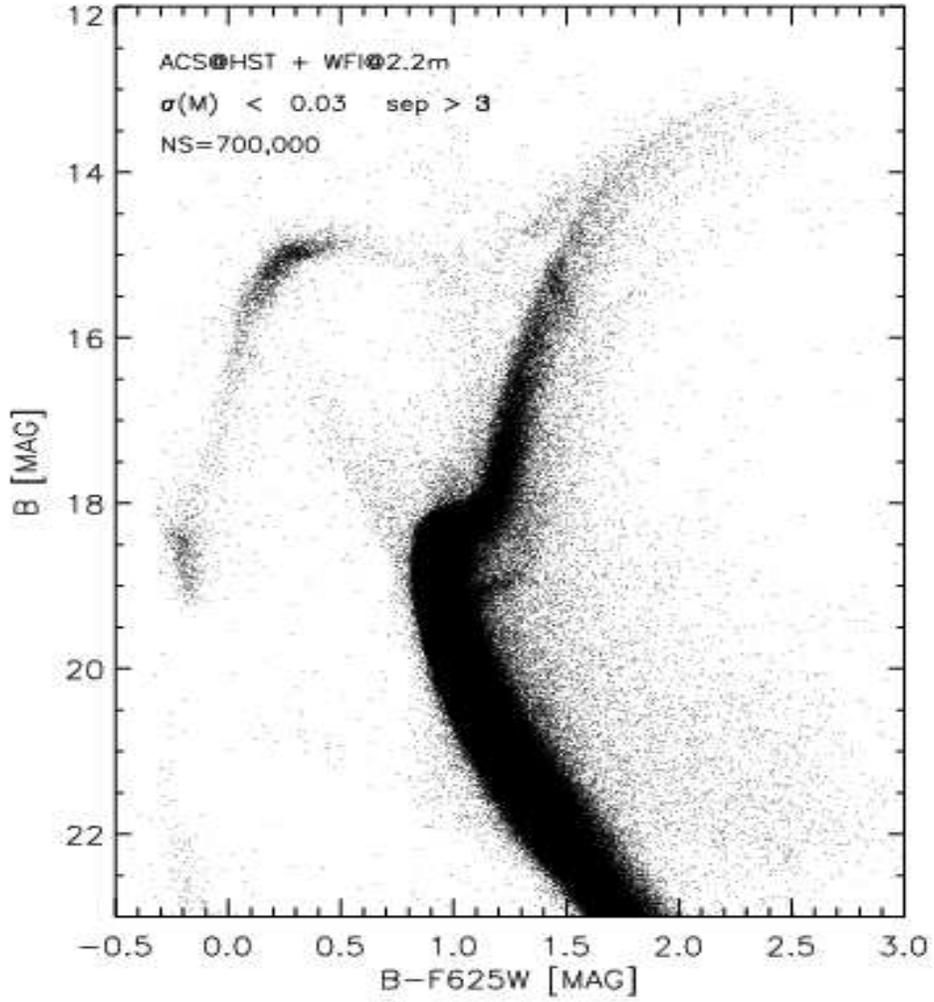}
\caption {
Composite $B, B-F625W$ diagram of the entire field. ACS data in the 
$F435W-$band  were transformed into the $B-$band, while $V$ and $I$ 
WFI data have been transformed into F625W using the transformation 
$F625W=V\times0.544 + I\times0.455$ mag. See text for more details. 
} \label{fig3}
\end{center}
\end{figure}

\clearpage
\begin{figure}[!h]
\begin{center}
\includegraphics[height=0.65\textheight,width=0.80\textwidth]{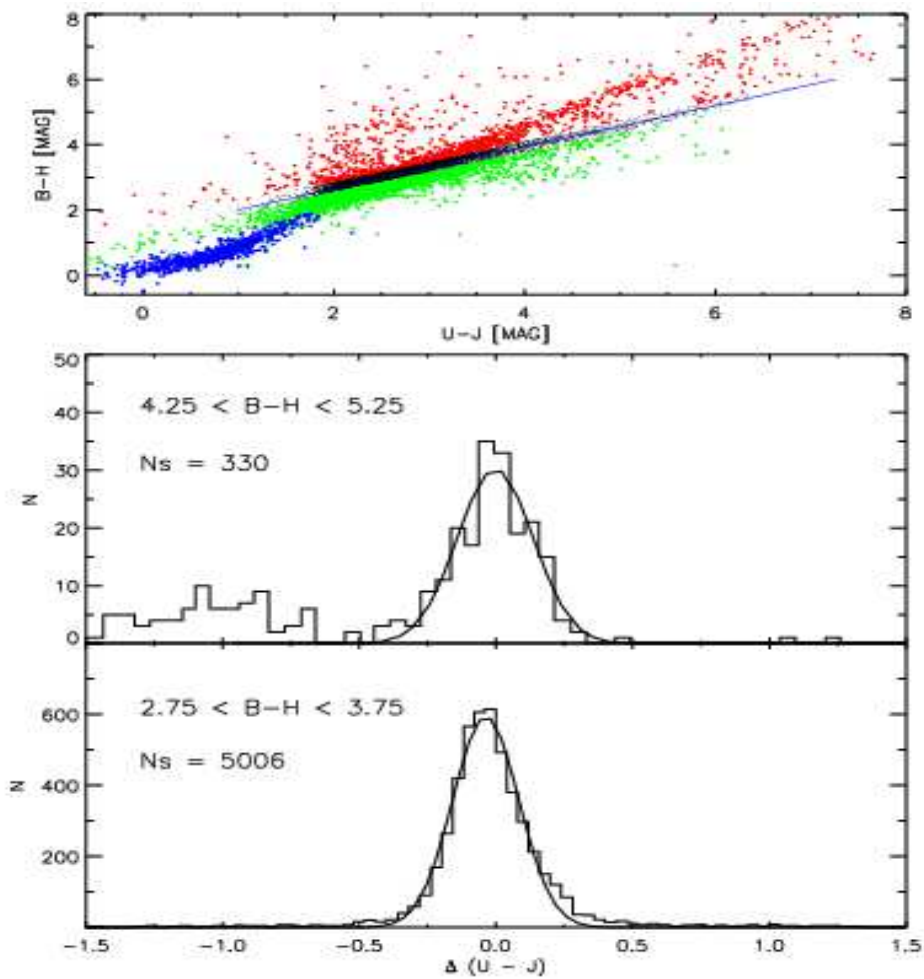}
\caption{Top: Optical-NIR color-color plane for bright stars in \omcp. Black and
blue dots display likely cluster RG and HB stars, while red and green 
dots likely field objects. The solid blue line shows the fitting line adopted to 
select likely field and cluster stars. Middle: distance in $U-J$ color from the 
fitting line for the stars with $4.25 \le B-H \le 5.25$ and located at a radial distance 
ranging from 2.5 to 10 arcmin from the cluster center. The solid line shows the gaussian 
that fits the color distribution. Bottom: Same as the middle, but for stars with 
$2.75 \le B-H \le 3.75$.       
}\label{fig4}
\end{center}
\end{figure}

\clearpage
\begin{figure}[!h]
\begin{center}
\includegraphics[height=0.65\textheight,width=0.80\textwidth]{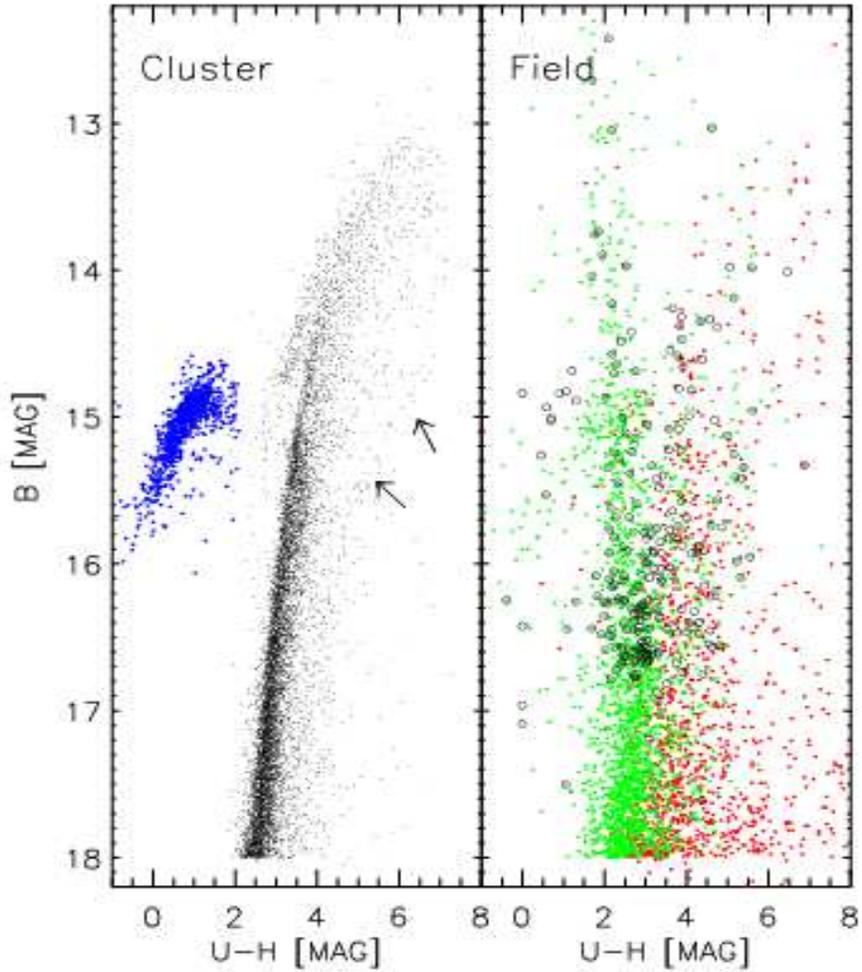}
\caption{Left: Optical-NIR CMD for bright likely cluster stars in \omcp. 
Different samples have the same colors as in Fig. 3. Right: same as left, but 
for likely field objects. Open circles mark objects that according to 
proper motion measurements (van Leeuwen et al. 2000) present a low membership 
probability ($P \le 10$\%).    
}\label{fig5}
\end{center}
\end{figure}

\clearpage
\begin{figure}[!h]
\begin{center}
\includegraphics[height=0.45\textheight,width=0.50\textwidth]{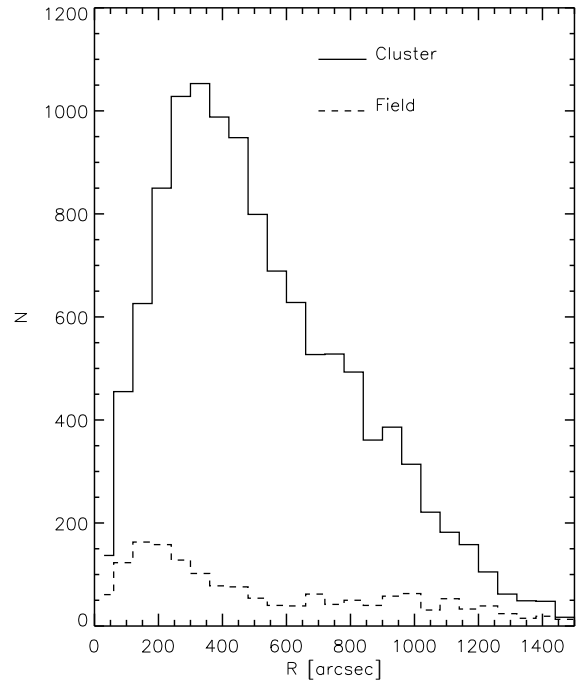}
\caption{Radial distribution of likely field objects (dashed line) and 
cluster (solid line) stars. 
}\label{fig6}
\end{center}
\end{figure}

\clearpage
\begin{figure}[!h]
\begin{center}
\includegraphics[height=0.55\textheight,width=0.77\textwidth,angle=180]{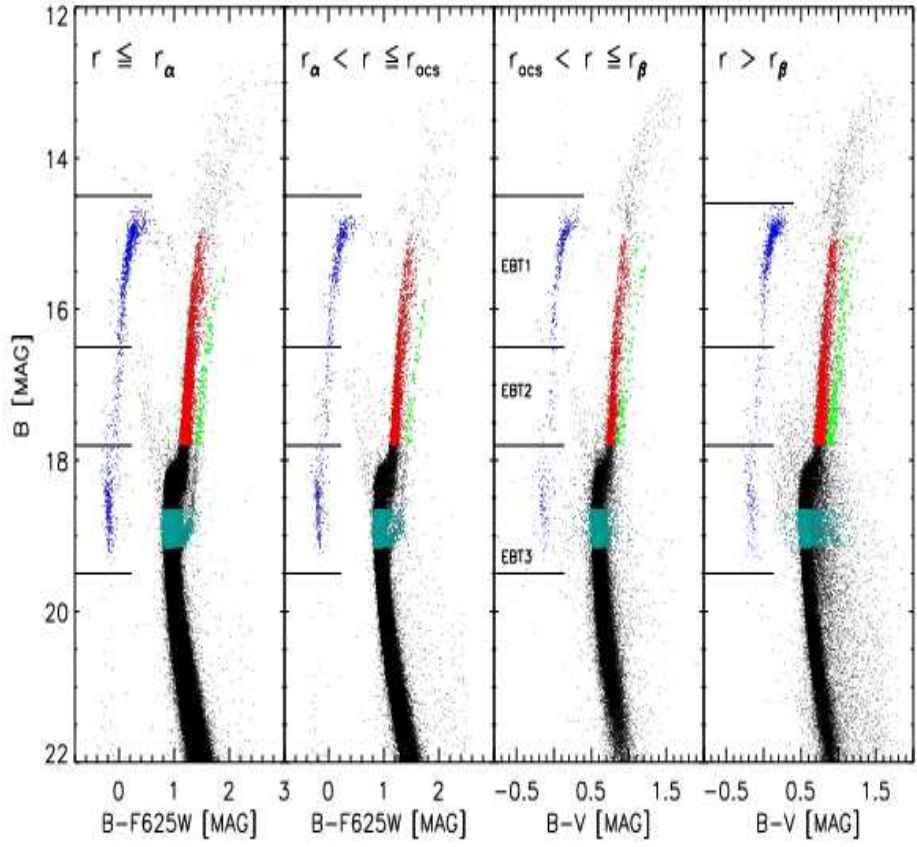}
\vspace*{0.15truecm}
\caption {
CMDs as a function of the radial distance based on ACS@HST ($B$, $B-F625W$) and 
ground-based ($B$, $B-V$) data. The bulk of the stars plotted in the different 
CMDs (black dots) were selected using different criteria ({\tt sep}, photometric 
errors, sharpness). Blue dots mark HB stars, red dots RGB stars, green dots 
stars belonging to the $\omega 3-$branch, and cyan dots MS stars adopted for the 
star counts. 
These sub-samples have only been selected between upper and lower magnitude
limits and cleaned for field star contamination.
The solid lines mark from top to bottom the EBT1, the EBT2, and the EBT3 region. 
\label{fig7}}
\end{center}
\end{figure}

\clearpage
\begin{figure}[!h]
\begin{center}
\includegraphics[height=0.65\textheight,width=0.80\textwidth]{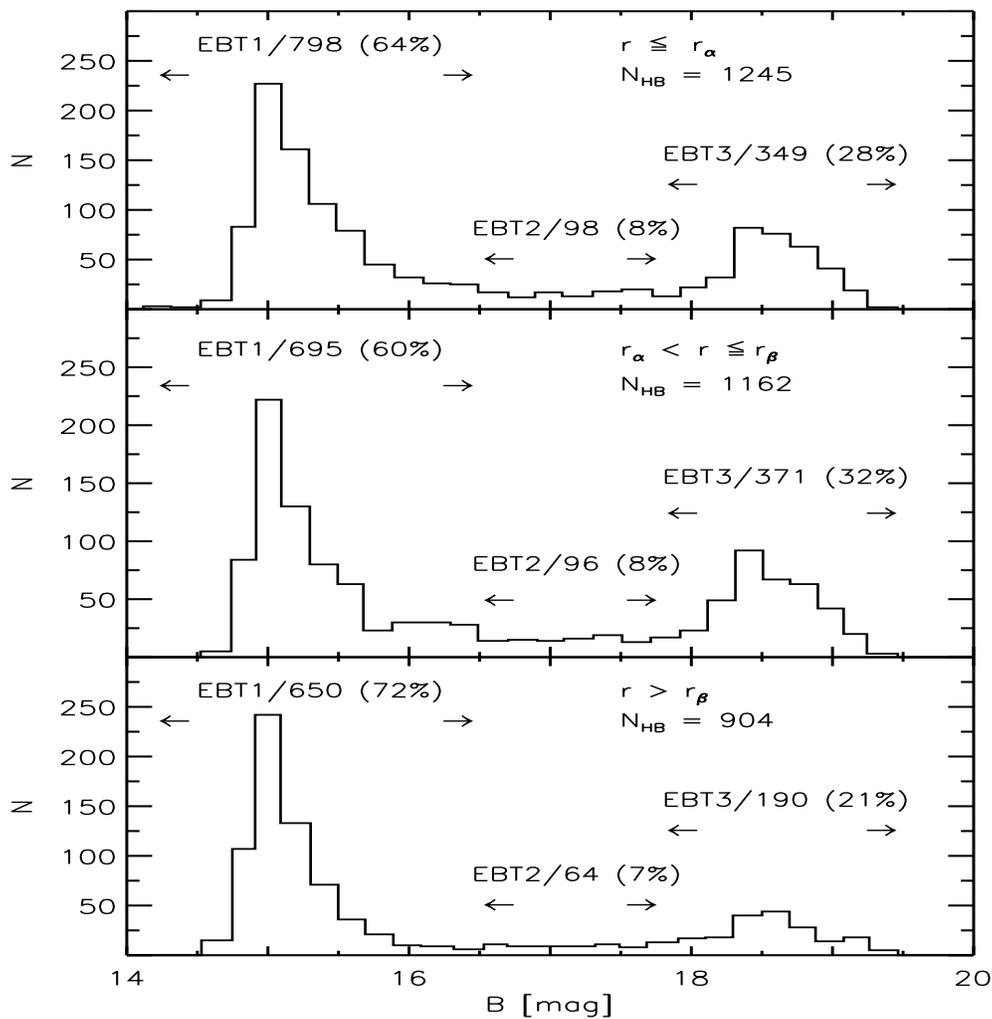}
\caption {
The $B-$band luminosity function of HB stars from the very center (top) to 
more external cluster regions (middle, bottom). The arrows mark the magnitude 
interval of the three selected EBT regions (see solid lines in Fig. 7). 
The total number of HB stars, the star counts for each sub-sample, and their 
relative fractions in each region are also labeled.  
}\label{fig8}
\end{center}
\end{figure}

\clearpage
\begin{figure}[!h]
\begin{center}
\includegraphics[height=0.55\textheight,width=0.77\textwidth,angle=180]{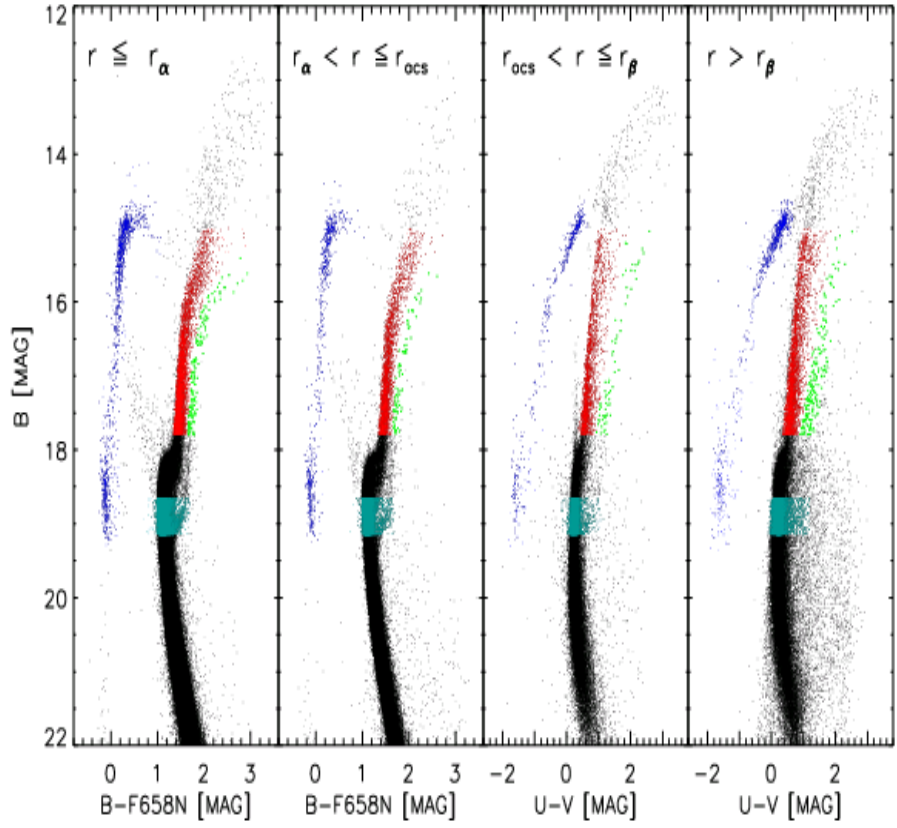}
\vspace*{0.10truecm}
\caption{Same as Fig. 7, but based on ACS@HST ($B$, $B-F658N$) and ground-based 
WFI@2.2m ESO/MPI ($B$, $U-V$) data. 
} \label{fig9}
\end{center}
\end{figure}

\clearpage
\begin{figure}[!h]
\begin{center}
\includegraphics[height=0.65\textheight,width=0.80\textwidth]{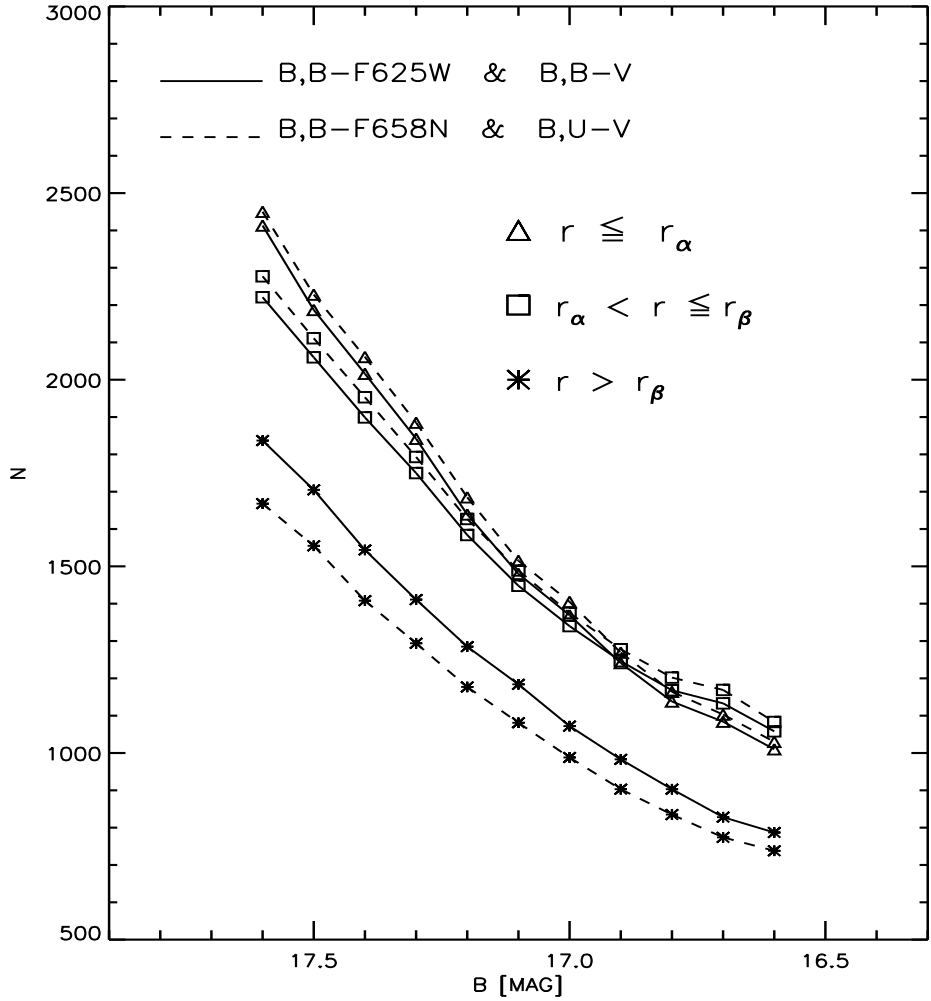}
\caption{Empirical star counts along the RGB of $\omega$ Cen. Individual 
bins cover an interval of one $B-$magnitude and have been plotted at the 
faint end of the magnitude bin. Different symbols mark star counts in 
different annuli. Solid and dashed lines show RG samples selected in 
different CMDs.   
} \label{fig10}
\end{center}
\end{figure}

\clearpage
\begin{figure}[!h]
\begin{center}
\includegraphics[height=0.65\textheight,width=0.80\textwidth]{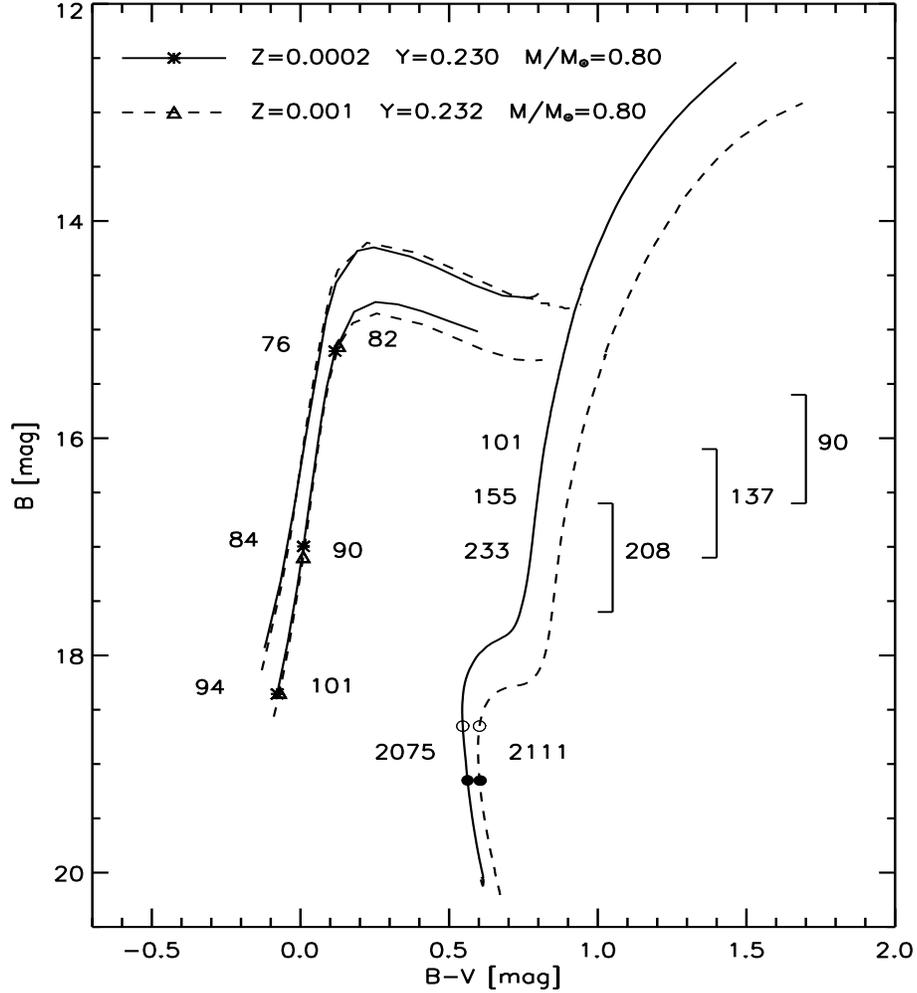}
\caption{Evolutionary predictions in the $B,B-V$ CMD at fixed mass 
($M=0.80 M_\odot$) and different 
chemical compositions (see labeled values). The square brackets along the 
evolutionary tracks mark three out of the eleven one magnitude bins along 
the RGB. The 
labels display the predicted lifetimes (Myr) for both metal-poor (left, 
Z=0.0002) and more metal-rich (right, Z=0.001) structures. 
Solid and dashed lines for He burning structures show the ZAHB and the 
exhaustion of central He burning with different chemical compositions.  
The labels and the symbols along the ZAHBs display predicted lifetimes 
(Myr) of metal-poor (left, asterisks) and more metal-rich (right, triangles) 
He burning structures representative from top to bottom of EBT1, EBT2, 
and EBT3 groups. Solid and empty circles mark the region of the MS adopted 
to estimate the MS lifetime.} 
\label{fig11}
\end{center}
\end{figure}

\clearpage
\begin{figure}[!h]
\begin{center}
\includegraphics[height=0.65\textheight,width=0.80\textwidth]{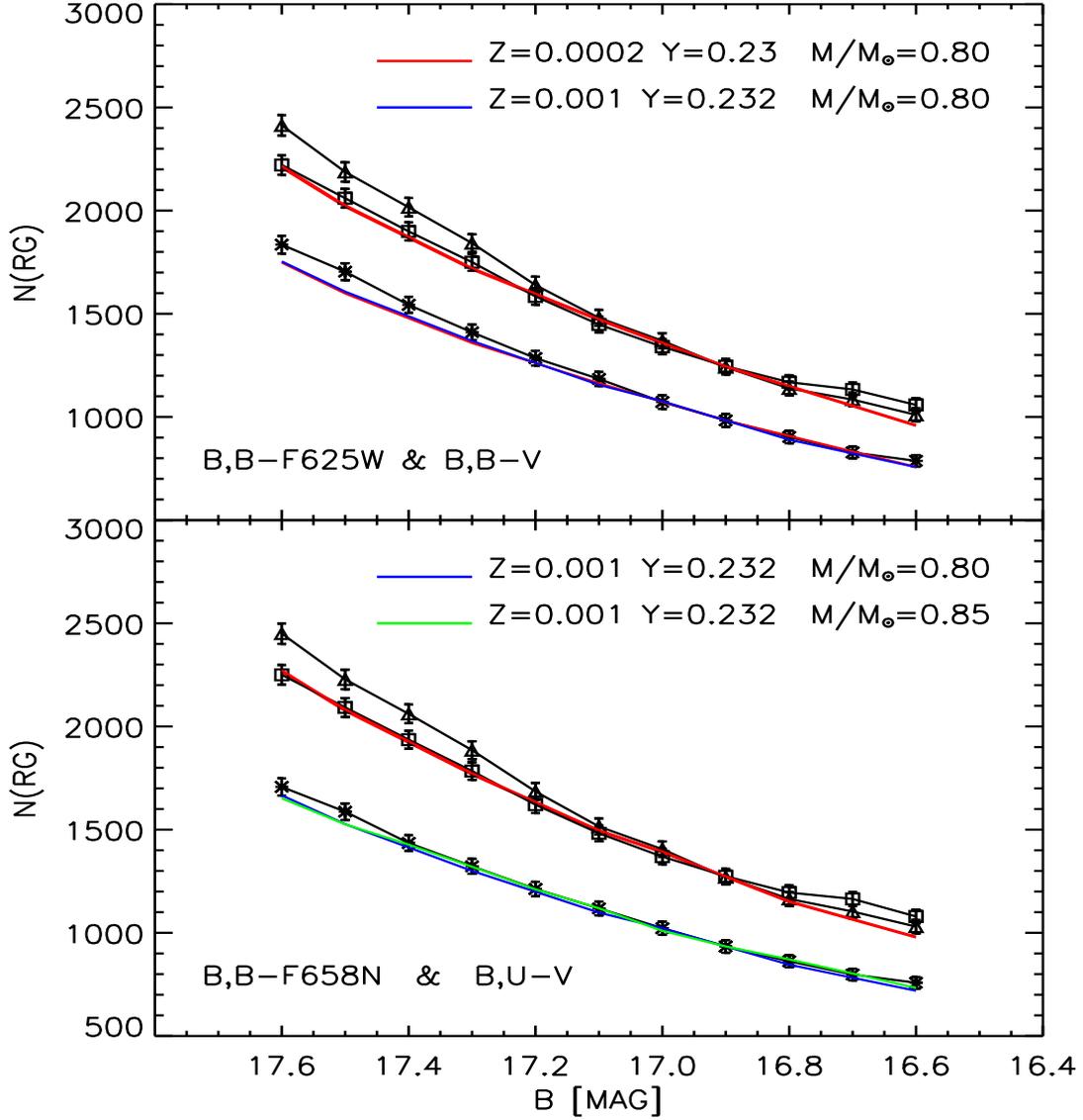}
\vspace*{0.75truecm}
\caption{Top: Comparison between empirical star counts along the RGB of 
$\omega$ Cen and predicted lifetimes for different chemical compositions 
and fixed stellar mass (see labeled values). Theoretical predictions have 
been normalized to observations in the magnitude bin $m_B=16.9-15.9$ 
($M_B=2.74-1.75$). 
The normalization was separately applied to the star counts of the three
different annuli.
The symbols are the same as in Fig. 10 and the error 
bars only account for uncertainties (Poisson) on star counts. 
Bottom: Same as the top, but for RG stars selected in the CMDs 
$B,B-F652N$ (ACS); $B,U-V$ (WFI). Predicted lifetimes refer to stellar 
structures constructed at fixed chemical composition and different stellar 
masses (see labeled values).  
} \label{fig12}
\end{center}
\end{figure}

\clearpage
\begin{figure}[!h]
\begin{center}
\includegraphics[height=0.65\textheight,width=0.80\textwidth]{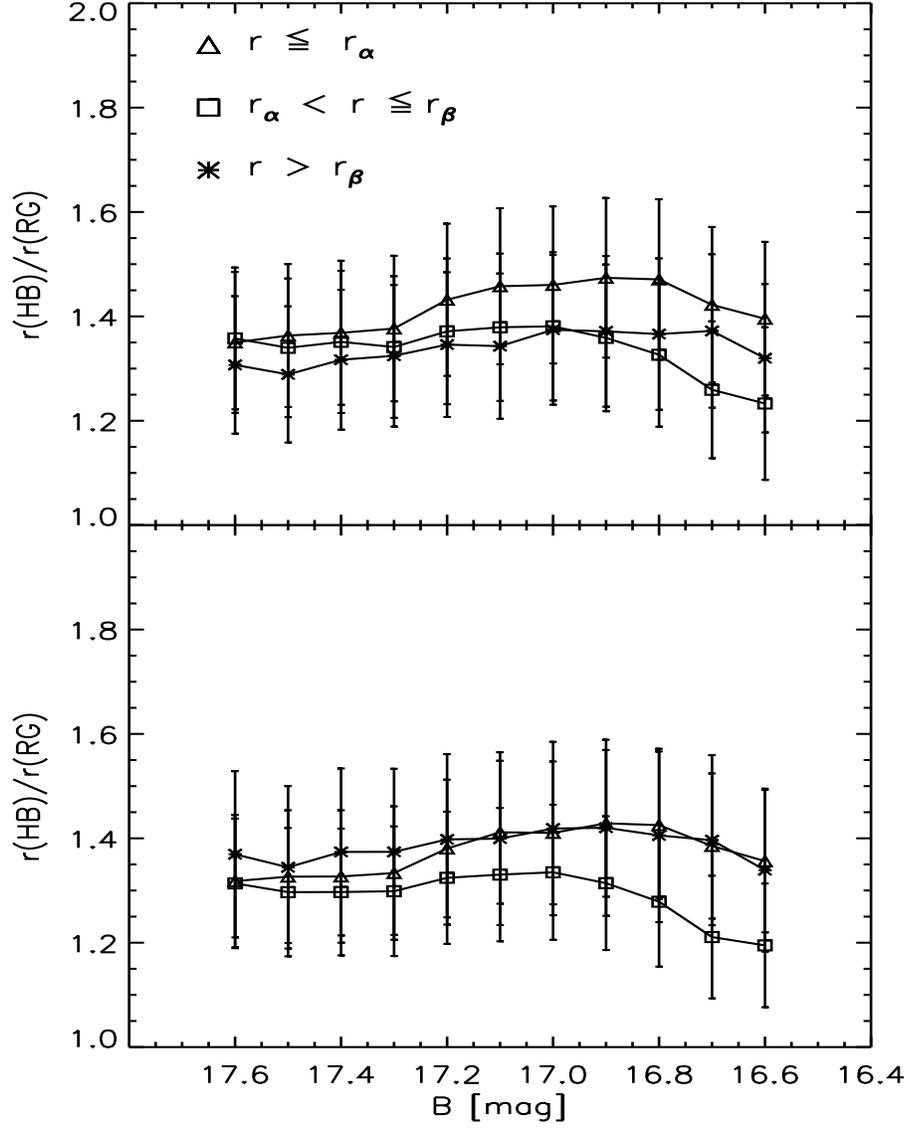}
\vspace*{0.75truecm}
\caption{Top: Ratios between the rate of HB stars and the rate of RG stars in 
the three different annuli. Star counts are based on $B,B-F625W$ (ACS) and 
$B,B-V$ (WFI) CMDs. The symbols are the same as in Fig. 10 and the error bars 
account for uncertainties on star counts (Poisson) and on evolutionary lifetimes (10\%). 
Bottom: Same as the top, but for star counts based on $B,B-F658N$ (ACS) and 
$B,U-V$ (WFI) CMDs. 
\label{fig13}} 
\end{center}
\end{figure}

\clearpage
\begin{figure}[!h]
\begin{center}
\includegraphics[height=0.65\textheight,width=0.80\textwidth]{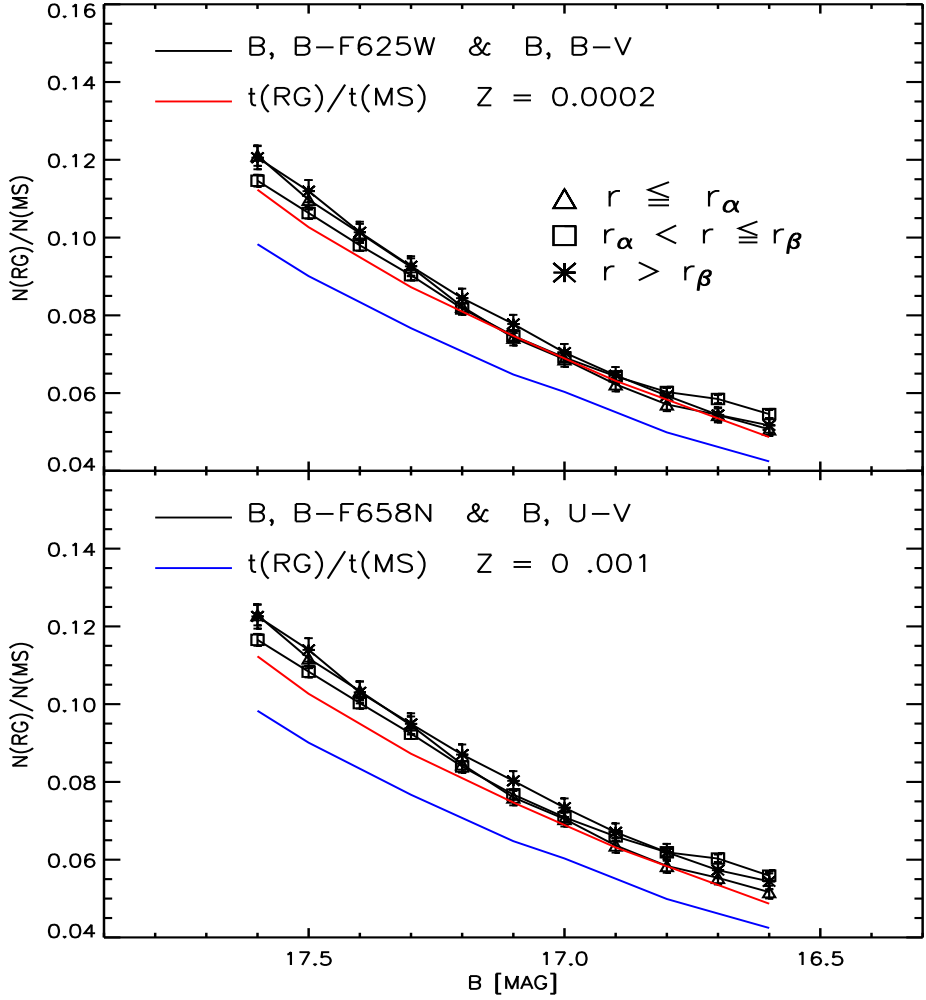}
\vspace*{0.65truecm}
\caption{Top: Ratio between RG and MS star counts selected in the $B,B-F625W$ 
(ACS) and $B,B-V$ (WFI) CMDs. The red and the blue lines show the predicted ratio 
between RG and MS lifetimes at fixed stellar mass ($M= 0.80 M_\odot$) and 
different metal abundances (see labeled values). Theoretical predictions 
were not normalized to observed ratios. The symbols are the same as in Fig. 10 
and the error bars only account for uncertainties (Poisson) in star counts.    
Bottom: Same as the top, but the RG and MS star counts were selected in the 
$B,B-F658N$ (ACS) and $B,U-V$ (WFI) CMDs.   
\label{fig14}} 
\end{center}
\end{figure}

\clearpage
\begin{figure}[!h]
\begin{center}
\includegraphics[height=0.65\textheight,width=0.80\textwidth]{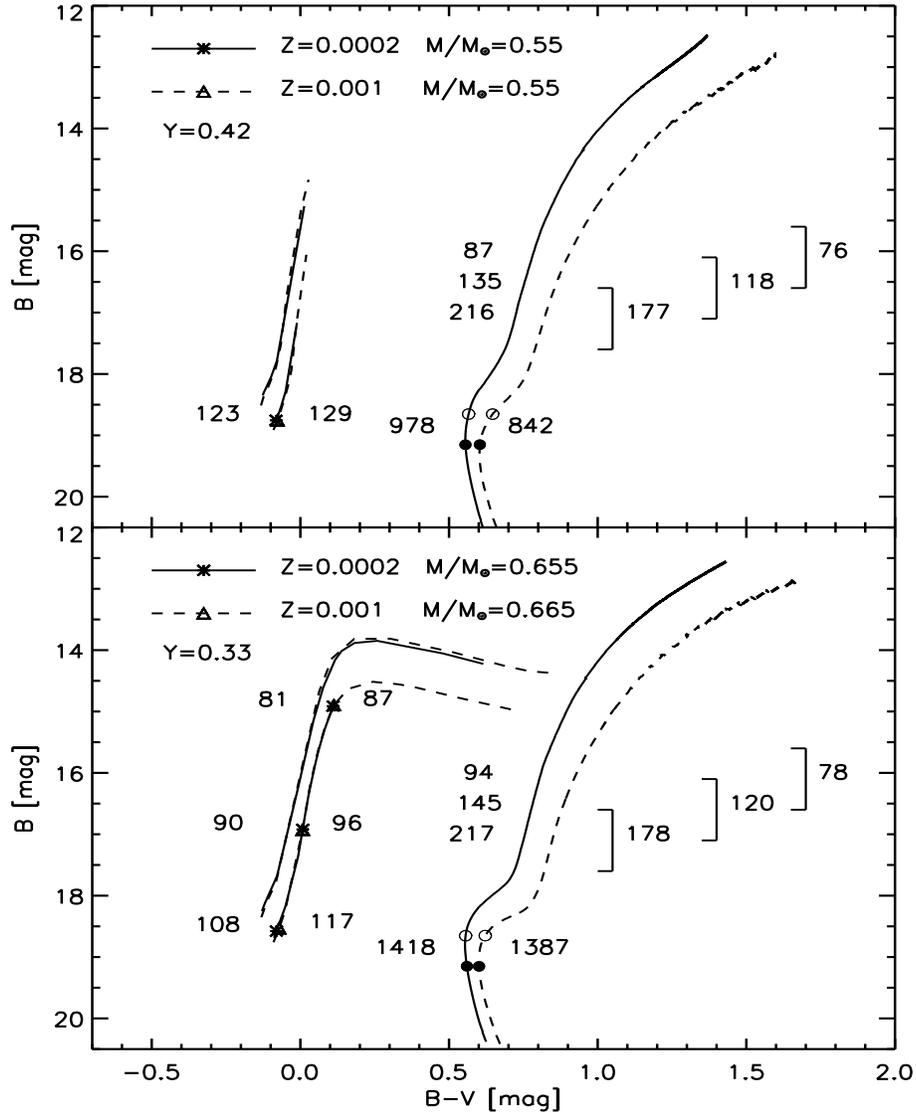}
\vspace*{0.65truecm}
\caption{Top: Same as Fig. 11. The evolutionary models refer to stellar 
structures constructed by adopting the same TO-age and metal abundances 
but a higher helium content (Y=0.42). Bottom: Same as the top, but for a 
lower helium content (Y=0.33).  
\label{fig15}} 
\end{center}
\end{figure}

\clearpage
\begin{figure}[!h]
\begin{center}
\includegraphics[height=0.65\textheight,width=0.80\textwidth]{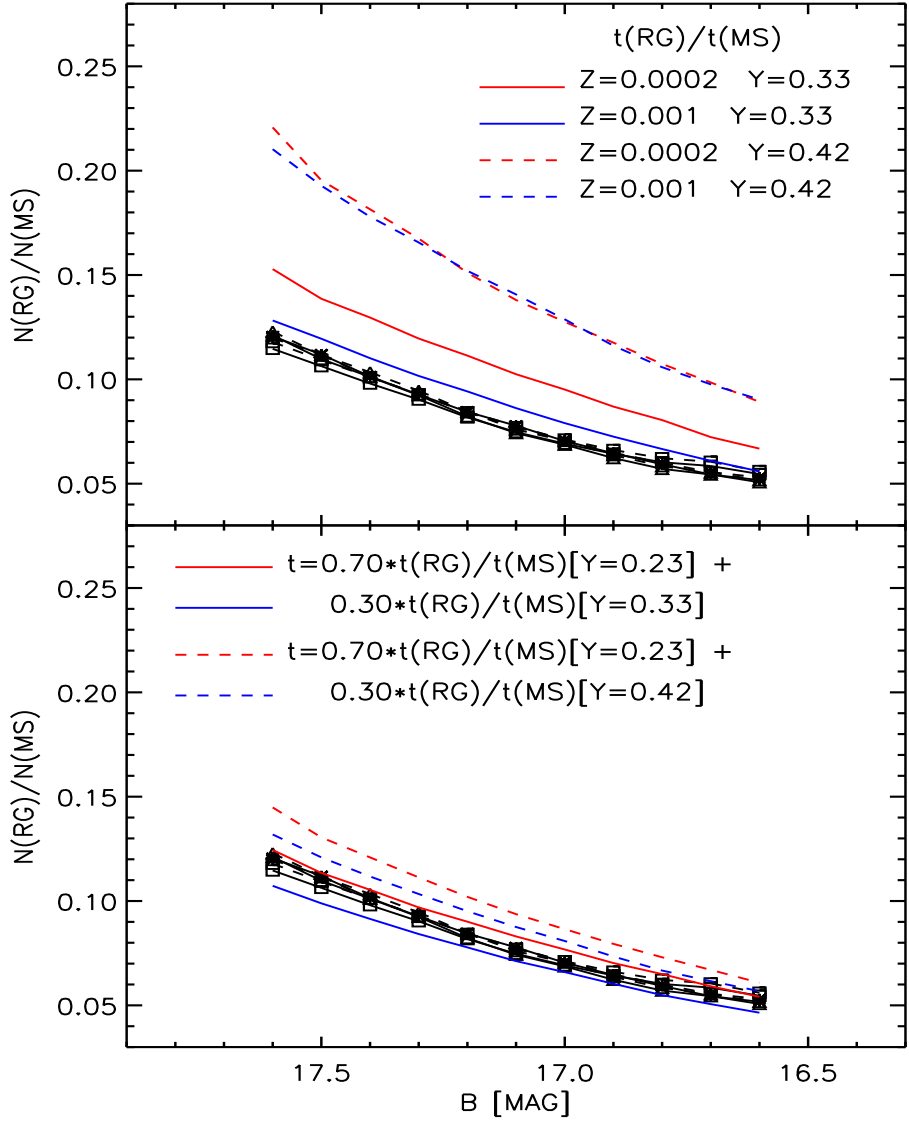}
\vspace*{0.65truecm}
\caption{Top: Same as Fig. 14. The theoretical ratios refer to the helium 
enhanced (Y=0.33, 0.42) stellar structures. Bottom: Same as the top but 
predicted ratios account for the fraction of He enhanced stellar population
(30\%). See text for more details.\label{fig16}} 
\end{center}
\end{figure}

\clearpage
\begin{figure}[!h]
\begin{center}
\includegraphics[height=0.65\textheight,width=0.80\textwidth]{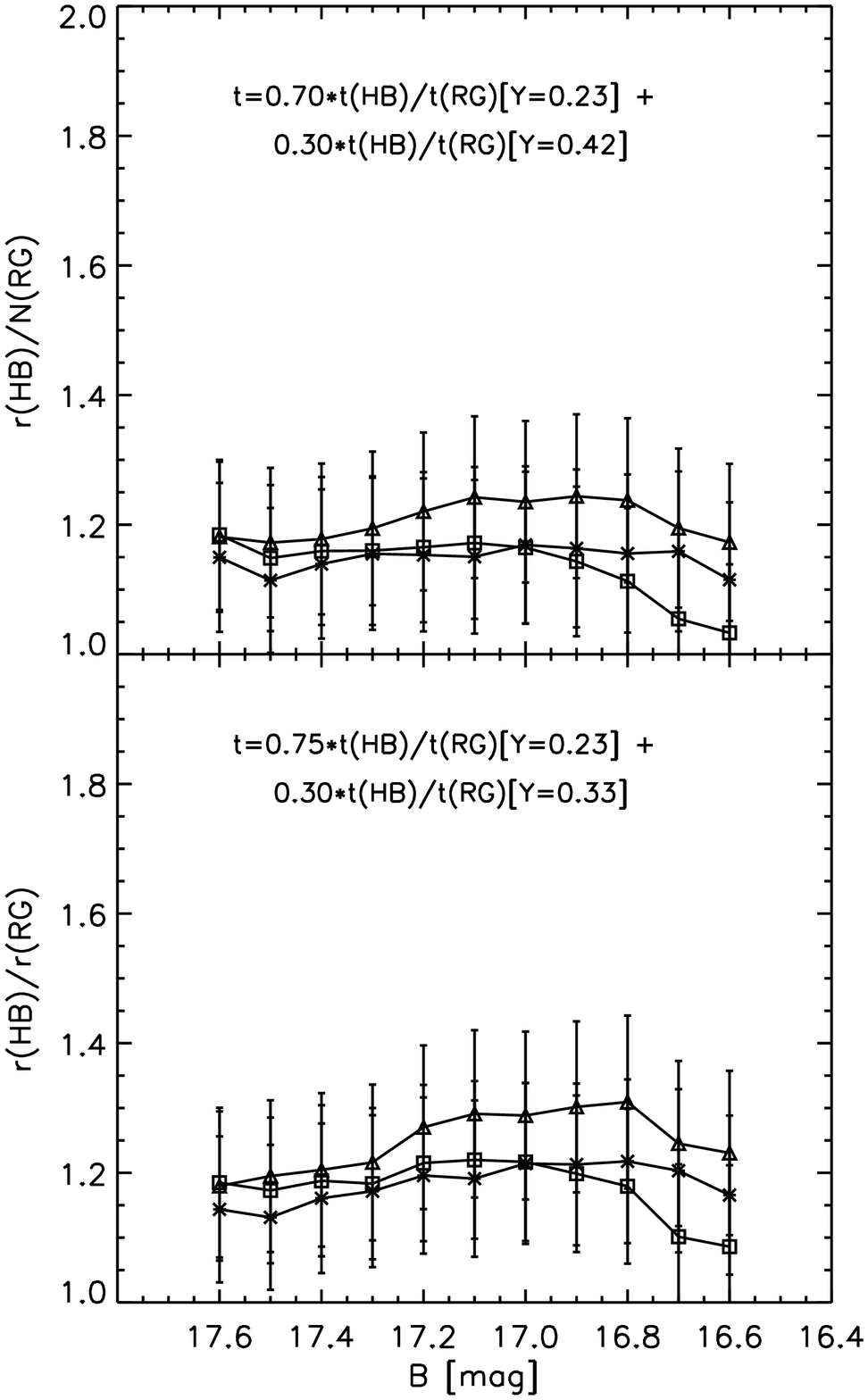}
\vspace*{0.65truecm}
\caption{Top: Comparison between the rates of HB and RG stars. 
Star counts are based on $B,B-F625W$ (ACS) and $B,B-V$ (WFI) CMDs. 
The symbols are the same as in Fig. 13 and the error bars account
for uncertainties on star counts (Poisson) and on evolutionary 
lifetimes (10\%). 
The predicted HB lifetimes account for stellar populations with 
different fractions of canonical (Y=0.23, 70\%) and He-enriched 
(Y=0.42, 30\%) stars and for the spread in metal abundance. 
Bottom: Same as the top, but for a He-enriched stellar population 
with Y=0.33.\label{fig17}} 
\end{center}
\end{figure}


\begin{thebibliography}{60}
\expandafter\ifx\csname natexlab\endcsname\relax\def\natexlab#1{#1}\fi
\bibitem[]{1555} Anderson, J. 2002, in {\it Omega Centauri: A Unique Window into 
Astrophysics}, ed. F. van Leeuwen, J. Hughes, \& G. Piotto (San Francisco: ASP), 
265, 87
\bibitem[]{1558} Baade, D. et al. 1999, The Messenger, 95, 15
\bibitem[]{1559} Bailyn, C.D., Sarajedini, A., Cohn, H., Lugger, P.M., \& Grindlay, J.E. 1992, \aj, 103, 1564
\bibitem[]{1560} Bedin, L. R., Piotto, G., Zoccali, M. et al. 2000, A\&A, 363, 159
\bibitem[]{1561} Bedin, L. R., Piotto, G., Anderson, J., Cassisi, S. et al. 2004, ApJ, 605, 125	
\bibitem[]{1735} Behr, B.B., Cohen, J.G., McCarthy, J.K. 2000, ApJ, 531, L37
\bibitem[]{1562} Bekki, K., \& Norris, J. E. 2006, ApJ, 637, L109 
\bibitem[]{1563} Bessel, M. S., Castelli, F. \& Plez, B. 1998, A\&A, 333, 231
\bibitem[]{1564} Bono G., Castellani V., Degl'Innocenti S., Pulone, L. 1995, A\&A 297, 115 
\bibitem[]{1565} Brown, T.M.,Sweigart, A.V., Lanz T., Landsman, W.B., \& Hubeny, I. 2001, \apj, 562, 368
\bibitem[]{1566} Calamida, A. et al. 2005, \apj, 634, 69
\bibitem[]{1567} Caloi, V., Castellani, V., \& Tornamb\`e, A. 1978, A\&AS, 33, 169
\bibitem[]{1742} Caputo, F., Chieffi, A., Tornambe, A., Castellani, V., Pulone, L. 1989, ApJ, 340,241
\bibitem[]{1568} Caputo F., Cassisi S. 2002, MNRAS, 333, 825
\bibitem[]{1569} Cardelli, J.A., Clayton, G.C., \& Mathis, J.S.  1989, \apj, 345, 245 
\bibitem[]{1570} Cariulo, P., Degl'Innocenti, S., \& Castellani, V. 2004, A\&A, 424, 927
\bibitem[]{1571} Cassisi, S., Castellani, V., Degl'Innocenti, S. \& Weiss, A. 1998, A\&AS, 129, 267
\bibitem[]{1572} Cassisi, S. et al. 2001, A\&A 366, 578
\bibitem[]{1573} Cassisi, S., Schlattl, H., Salaris, M., \& Weiss, A. 2003, \apj, L43
\bibitem[]{1576} Castellani, V., Chieffi, A., Pulone, L., \& Tornamb\'e, A. 1985, \apj, 296, 204
\bibitem[]{1574} Castellani, M., \& Castellani, V. 1993, ApJ, 407, 649 
\bibitem[]{1575} Castellani, M., Castellani, V., Pulone, L., \& Tornamb\`e, A. 1994, A\&A, 282, 771
\bibitem[]{1577} Castellani, V., Iannicola, G., Bono, G., Zoccali, M., Cassisi, S.,
 \& Buonanno, R. 2006a, A\&A, 446, 569
\bibitem[]{1578} Castellani, M., Castellani, V., \& Prada Moroni, P. G. 2006b, A\&A, 457, 569
\bibitem[]{1579} Castelli, F. \& Kurucz, R.L. 2003, in IAU Symposium 210, 
ed. N. Piskunov, W.W. Weiss, \& D.F. Gray,(Uppsala, Sweden), 20
\bibitem[]{1581} Chieffi, A., Straniero, O. 1989, \apjs, 71, 47
\bibitem[]{1582} Ciacio, F., Degl'Innocenti, S., Ricci, B. 1997, A\&AS, 123, 449
\bibitem[]{1583} Corsi, C.E. et al. 2003, MmSAI, 74, 884
\bibitem[]{1584} Cutri, R. M. \& 2MASS collaboration 2003, www.ipac.caltech.edu/2mass 
\bibitem[]{1585} Da Costa, G. S.,  Villumsen, J. V. 1981, in IAU. Coll. 68, 
{\it Astrophysical Parameters for Globular Clusters}, (Knudsen), 527
\bibitem[]{1587} Da Costa, G.S., Norris, J., Villumsen, J.V. 1986, \apj, 308, 743
\bibitem[]{1588} D'Antona, F., Bellazzini, M., Caloi, V., Fusi Pecci, F., Galleti, S., 
\& Rood, R. T. 2005, ApJ, 631, 868 
\bibitem[]{1766} Davies, M. B., Piotto, G., de Angeli, F. 2004, MNRAS, 349, 129
\bibitem[]{1590} D'Cruz, N.L., Dorman, B., Rood, R.T., O'Connel, R.W. 1996, \apj, 466, 359
\bibitem[]{1591} D'Cruz, N.L., Rood, R.T., O'Connel, R.W. et al. 2002,  in {\it Omega 
Centauri: A Unique Window into Astrophysics}, ed. F. van Leeuwen, J. Hughes, \& 
G. Piotto (San Francisco: ASP), 235
\bibitem[]{1594} Del Principe, M. et al. 2006, \apj, 77, 330
\bibitem[]{1772} Ferraro, F. R., Sollima, A., Rood, R. T., Origlia, L., Pancino, E., Bellazzini, M. 2006, ApJ, 638, 433
\bibitem[]{1773} Denissenkov, P. A., VandenBerg, Don A. 2003, ApJ, 598, 1246
\bibitem[]{1595} Freyhammer, L.M. et al. 2005, \apj, 623, 860
\bibitem[]{1596} Gilmore G., King I., van der Kruit P.C. 1990, in Saas-Fee advanced 
course, {\it The Milky Way as a galaxy}, ed. Buser R. \& King I., (Geneva), 19
\bibitem[]{1598} Harris, W.E. 1996, \aj, 112, 1487
\bibitem[]{1599} Hilker, M., Kayser, A., Richtler, T., \&  Willemsen, P. 2004, \aap, 422, L9
\bibitem[]{1600} Iben, I. Jr. 1991, \apjs, 76, 55 
\bibitem[]{1601} Iglesias, C.A., Rogers, F.J.  1996, \apj, 464, 943
\bibitem[]{1602}  Ivanova, N., Heinke, C. O., Rasio, F. A., Taam, R. E., 
Belczynski, K., \& Fregeau, J. 2006, MNRAS, 372, 1043 
\bibitem[]{1604} Jordi, K., Grebel, E. K., \&  Ammon, K. 2006, A\&A, accepted, astro-ph/0609121 
\bibitem[]{1605} Kaluzny, J., Thompson, I., Krzeminski, W., Olech, A., Pych, W., 
\& Mochejska, B. 2002,  in {\it Omega Centauri: A Unique Window into Astrophysics}, 
ed. F. van Leeuwen, J. Hughes, \& G. Piotto (San Francisco: ASP), 155
\bibitem[]{1608} King, I. 1962, AJ, 67, 471
\bibitem[]{1788} King, A. R., Surdin, V. G., \& Rastorguev, A. S. 2002, Vvedenie v Klassiøceskuju Zvezdnuju Dinamiku (Moskva: Editorial URSS)
\bibitem[]{1609} Lee, Y.-W. et al. 2005, \aj, 621, L57
\bibitem[]{1610} Lub, J. 2002, in {\it Omega Centauri: A Unique Window into
Astrophysics}, ed. F. van Leeuwen, J. Hughes, \& G. Piotto (San Francisco: ASP),  95
\bibitem[]{1612} Maeder, A., \& Meynet, G.  2006, A\&A, 448, L37 
\bibitem[]{1793} Melbourne, J., Sarajedini, A., Layden, A., Martins, D. H. 2000, AJ, 120, 3127   
\bibitem[]{1613} Meylan, G., Mayor, M., Duquennoy, A., \& Dubath, P. 1995, A\&A, 303, 761
\bibitem[]{1614} Moehler, S., Sweigart, A. V., Landsman, W. B., Dreizler, S. 2002, A\&A, 395, 37
\bibitem[]{1614} Moehler, S., \& Sweigart, A. V. 2006, A\&A, 455, 943 
\bibitem[]{1615} Momany, Y., et al. 2004, A\&A, 420, 605
\bibitem[]{1616} Monelli, M. et al. 2003, \aj, 126, 218
\bibitem[]{1617} Monelli, M., Corsi, C.E., Castellani, V. et al. 2005, \apj, 621, L117
\bibitem[]{1618} Norris, J. E.  2004, ApJ, 612, 25
\bibitem[]{1618} Norris, J. E., Freeman, K. C., Mayor, M., \& Seitzer, P. 1997, ApJ, 487, L187
\bibitem[]{1618} Norris, J. E., Freeman, K. C., \& Mighell, K. J. 1996, \apj, 462, 241
\bibitem[]{1619} Ochsenbein, F., Bauer, P., \& Marcout, J. 2000, A\&AS, 143, 221
\bibitem[]{1804} Palacios, A., Charbonnel, C., Talon, S., Siess, L. 2006, A\&A, 453, 261
\bibitem[]{1620} Pancino, E., Ferraro, F. R., Bellazzini, M., Piotto, G., Zoccali, M. 2000, \apj, 534, 83
\bibitem[]{1621} Pancino, E., Pasquini, L., Hill, V., Ferraro, F.R., \& Bellazzini, M. 2002, ApJL, 568, L101
\bibitem[]{1622} Pietrinferni, A., Cassisi, S., Salaris, M. and Castelli, F. 2004, \apj, 612, 168
\bibitem[]{1623} Pietrinferni, A., Cassisi, S., Salaris, M. and Castelli, F. 2006, \apj, 642, 697
\bibitem[]{1809} Piotto, G., et al. 2004, \apj, 604, L109
\bibitem[]{1624} Piotto, G. et al. 2005, \apj, 621, 777
\bibitem[]{1625} Pollard, D. L., Sandquist, E. L., Hargis, J. R., Bolte, M. 2005, ApJ, 
628, 729 
\bibitem[]{1813} Recio-Blanco, A., Aparicio, A., Piotto, G., de Angeli, F., Djorgovski, S. G 2006, A\&A, 452, 875
\bibitem[]{1814} Recio-Blanco, A., Piotto, G., Aparicio, A., Renzini, A. 2004, A\&A, 417, 597
\bibitem[]{1815} Renzini, A. 1977, in {\it The Evolution of Population II Stars and Mass 
Loss and Stellar Evolution}, ed. P. Bouvier and A. Maeder, (Sauverny, Switzerland), 151
\bibitem[]{1627} Rey, S-C. et al. 2000, \aj, 119, 1824
\bibitem[]{1628} Rey, S-C., Joo, J-M., Sohn, Y-J., Ree C-H., Lee, Y-W. 2002, in 
{\it Omega Centauri: A Unique Window into Astrophysics}, ed. F. van Leeuwen, 
J. Hughes, \& G. Piotto (San Francisco: ASP), 177
\bibitem[]{1628} Romano, D., et al. 2007, MNRAS, accepted, astro-ph/0701162
\bibitem[]{1631} Salaris, M., Weiss, A., Ferguson, J. W., \& Fusilier, D. J. 2006, ApJ, 645, 1131
\bibitem[]{1632} Salaris, M.; Riello, M.; Cassisi, S.; Piotto, G. 2004, A\&A, 420, 911 
\bibitem[]{1633} Sandquist, E. L., Bolte, M., Langer, G. E., Hesser, J. E., \& Mendes de Oliveira, C. 1999, ApJ, 518, 262 
\bibitem[]{1634} Sandquist, E. L. \& Martel, A. R. 2006, ApJ Letters, accepted, astro-ph/0611278
\bibitem[]{1635} Sirianni, M., Jee, M. J., Benatez, N., et al. 2005, PASP, 117, 1049
\bibitem[]{1636} Sollima, A., Pancino, E., Ferraro, F. R., Bellazzini, M., Straniero, O., \& Pasquini, L. 2005, \apj, 634, 332
\bibitem[]{1637} Sollima, A., Borissova, J., Catelan, M., Smith, H. A., Minniti, D., 
Cacciari, C., \& Ferraro, F.R. 2006, AJ, 640, L43
\bibitem[]{1638} Sollima, A., Ferraro, F. R., Bellazzini, M., Origlia, L., Straniero, O., \& Pancino, E., 2007, ApJ, 654, 915 
\bibitem[]{1640}  Spergel, D. N. WMAP collaboration, ApJ, submitted, astro-ph/0603449
\bibitem[]{1641} Stetson, P. B. 1991, in {\it The formation and evolution of star 
clusters}, ed. K. Janes, (San Francisco: ASP), 88   
\bibitem[]{1643} Stetson, P. B. 2000, PASP, 112, 925
\bibitem[]{1644} Stetson, P. B., Bruntt, H., \& Grundahl, F. 2003, \pasp, 115, 413 
\bibitem[]{1836} Straniero, O., Dominguez, I., Imbriani, G., Piersanti, L.  2003, ApJ, 583, 878
\bibitem[]{1837} Suntzeff, N. B., \& Kraft, R. P. 1996, \aj, 111, 1913
\bibitem[]{1645} Sweigart, A. V. 1997, in Third Conference on Faint Blue Stars, ed. A.G.D. Phillip, J. Liebert, \& R.A. Saffer (Schenrctaday: L.Davis Press), 3 
\bibitem[]{1839} Sweigart, A. V., Catelan, M. 1998, \apj, 501, L63
\bibitem[]{1646} Thoul, A., Bahcall, J., Loeb, A. 1994, \apj, 421, 828
\bibitem[]{1647} Trager, S. C., King, I. R., \& Djorgovski, S. 1995, AJ, 109, 218 
\bibitem[]{1648} van de Ven, G., van den Bosch, R. C. E., Verolme, E. K., \&
de Zeeuw, P. T. 2006, A\&A, 445, 513
\bibitem[]{1650} van Leeuwen, F., Le Poole, R. S., Reijns, R. A.,
Freeman, K. C., \& de Zeeuw, P. T. 2000, A\&A, 360, 472
\bibitem[]{1652} Walker, A. R. 1994, \pasp, 106, 828
\bibitem[]{1653} Weldrake, D. T. F., Sackett, P. D. \& Bridges, T. J. 2006, AJ, 
accepted, astro-ph/0610704
\bibitem[]{1655} Whitney, J.H., O'Connell, R.W., Rood, R.T. et al. 1994, \aj, 108, 1350
\bibitem[]{1656} Whitney, J.H., Rood R.T., O'Connell R.W et al. 1998, \apj, 495, 284
\bibitem[]{1657} Yasuda, N., \& SDSS collaboration 2001, \aj, 122, 1104  
\bibitem[]{1658} Zoccali, M., Cassisi, S., Bono, G., Piotto, G., Rich, R. M., 
\& Djorgovski, S. G. 2000, ApJ, 538, 289
\bibitem[]{1660} Zoccali, M., \& Piotto, G. 2000, A\&A, 358, 943
\end{thebibliography}
\end{document}